\documentclass[journal,twoside,doublecolumn]{IEEEtran}

\IEEEoverridecommandlockouts

\usepackage{cite}
\usepackage{ieeefig}
\usepackage{amsfonts}
\usepackage{epsfig,color}
\usepackage{graphicx}
\usepackage[caption=false,font=footnotesize]{subfig}
\usepackage{stfloats}
\usepackage{amsmath}
\usepackage{paralist}

\newtheorem{theorem}{\bf Theorem}
\newtheorem{lemma}{\bf Lemma}
\newtheorem{corollary}{\bf Corollary}

\newtheorem{definition}{\bf Definition}

\newcommand{\define}    {\stackrel{\triangle}{=}}  
\newcommand{\mg}{\rm{MG}}
\newcommand{\act}{asymmetrically constrained transmitters}

\begin{document}
%
\title{\Huge{A New Outer-Bound via Interference Localization and the Degrees of Freedom Regions of MIMO Interference Networks with no CSIT} }

\author{Chinmay S.~Vaze and Mahesh K.~Varanasi
\thanks{
The authors are with the Department of Electrical, Computer, and
Energy Engineering, University of Colorado, Boulder, CO 80309-0425
USA (e-mail: {vaze, varanasi}@colorado.edu).} }

%

\markboth{\MakeLowercase{Submitted,} IEEE T\MakeLowercase{rans}. I\MakeLowercase{nform}. T\MakeLowercase{h.}, M\MakeLowercase{ay}. 2011}{C. S. V\MakeLowercase{aze and} M. K. V\MakeLowercase{aranasi}: I\MakeLowercase{nterference} L\MakeLowercase{ocalization and the} D\MakeLowercase{o}F R\MakeLowercase{egions of the} MIMO IC \MakeLowercase{and} CRC \MakeLowercase{with} n\MakeLowercase{o} CSIT}


\maketitle


\begin{abstract}
The two-user multi-input, multi-output (MIMO) interference and cognitive radio channels are studied under the assumption of no channel state information at the transmitter (CSIT) from the degrees of freedom (DoF) region perspective. With $M_i$ and $N_i$ denoting the number of antennas at transmitter $i$ and receiver $i$ respectively, the DoF regions of the MIMO interference channel were recently characterized by Huang et al., Zhu and Guo, and by the authors of this paper for all values of numbers of antennas except when $\min(M_1,N_1) > N_2 > M_2$ (or $\min(M_2,N_2) > N_1 > M_1$). This latter case was solved more recently by Zhu and Guo who provided a tight outer-bound. Here, a simpler and more widely applicable proof of that outer-bound is given based on the idea of {\em interference localization}. Using it, the DoF region is also established for the class of MIMO cognitive radio channels when $\min(M_1+M_2,N_1) > N_2 > M_2$ (with the second transmitter cognitive) -- the only class for which the inner and outer bounds previously obtained by the authors were not tight -- thereby completing the DoF region characterization of the general 2-user MIMO cognitive radio channel as well.

\end{abstract}
%



\section{Introduction}
\IEEEPARstart{C}{onsider} a multiple-input multiple-output (MIMO) interference channel (IC) consisting of two transmitters, T1 and T2, equipped with $M_1$ and $M_2$ antennas, respectively, and their paired or intended receivers R1 and R2 having $N_1$ and $N_2$ antennas, respectively. Each transmitter must communicate its message to its paired receiver over a shared additive Gaussian noise channel so that its transmission produces interference at the unpaired receiver. Denote such a channel as the $(M_1,M_2,N_1,N_2)$ MIMO IC. The input-output relationship in this MIMO IC
is given as
\begin{eqnarray}
\mbox{R1: } \hspace{1pt} Y(t) = H^{11}(t) X^1(t) + H^{12}(t) X^2(t) + W(t),
\label{eq: IC model missing case R1} \\
\mbox{R2: } Z(t) = H^{21}(t) X^1(t) + H^{22}(t) X^2(t) + W'(t), \label{eq: IC model missing case R2}
\end{eqnarray}
where at the $t^{th}$ channel use, $Y(t) \in \mathbb{C}^{N_1 \times 1}$ and $Z(t)\in \mathbb{C}^{N_2 \times 1} $ are the signals received by R1 and R2, respectively; $X^1(t) \in \mathbb{C}^{M_1 \times 1}$ and $X^2(t) \in \mathbb{C}^{M_2 \times 1}$ are the signals transmitted by T1 and T2, respectively; $W(t)$ and $W'(t)$ are the additive white Gaussian noises; $H^{ij}(t) \in \mathbb{C}^{N_i \times M_j}$ represents the channel matrix between Tj and Ri, $i,j \in \{1,2\}$; there is a power constraint of $P$ at both transmitters, i.e.,
\[
\lim_{b \to \infty} \frac{1}{b} \sum_{t=1}^b  \mathbb{E} ||X_i(t)||^2 \leq P, ~ i=1,2.
\]

Recently, \cite{Chiachi2, D.Guo, Vaze_Dof_final} studied the DoF region of the MIMO IC with CSIR (i.e., with receivers having perfect channel knowledge) but with no CSIT. They provided inner and outer-bounds to the DoF region which coincide for a large class of MIMO ICs. In particular, these bounds yield the exact characterization of the no-CSIT DoF region except if either of the two inequalities, namely, $\min(M_1,N_1) > N_2 > M_2$ or its symmetric counterpart\footnote{Henceforth, we restrict attention to ICs with $\min(M_1,N_1) > N_2 > M_2$ without loss of generality.}, namely  $\min(M_2,N_2) > N_1 > M_1$, holds. For this latter class, \cite{Zhu_Guo_noCSIT_DoF_2010} more recently obtained a tight outer-bound and proved that the inner-bound proposed earlier in \cite{Chiachi2, D.Guo, Vaze_Dof_final} is indeed equal to the DoF region. The DoF region of the MIMO IC was determined earlier under the idealized CSIT (and CSIR) assumption in \cite{Chiachi-Jafar, Jafar-Maralle}.

Henceforth, MIMO ICs of interest in this work for which $\min(M_1,N_1) > N_2 > M_2$ will be referred to as having {\em asymmetrically constrained transmitters}. In the following, we describe briefly why the outer-bounds of \cite{Chiachi2, D.Guo, Vaze_Dof_final} are not tight for such MIMO ICs. Suppose $d_2$ DoF are to be achieved for the second (T2-R2) pair. Then, Fano's inequality \cite{CT} can be used to show that the total interference at R2 can not have a multiplexing gain higher than $N_2 - d_2$. Since the interference at R2 is caused by the transmission of T1, this condition puts constraints on $X^1(t)$, and hence on $d_1$ (the DoF of first pair T1-R1). Indeed, an outer-bound derived based on this idea suffices to characterize the no-CSIT DoF regions of all MIMO ICs except those with asymmetrically constrained transmitters. In this latter case, since $M_2 < N_2$, the transmit signal of T2, namely, $X^2(t)$ can not span the entire $N_2$-dimensional receive-signal space of R2. Thus, if $d_2$ DoF are to be achieved for the second pair, the interference at R2 should satisfy not just the constraint that its multiplexing gain can not exceed $N_2 - d_2$, but also that there exists an $M_2$-dimensional subspace at R2 which carries interference whose multiplexing gain is not more than $M_2 - d_2$; because if any $M_2$-dimensional subspace at R2 contains interference with multiplexing gain (strictly) more than $M_2 - d_2$, then R2 can not achieve $d_2$ DoF by decoding $X^2(t)$ since $X^2(t)$ lies within just an $M_2$-dimensional subspace. Accounting for this latter constraint becomes crucial for an IC with asymmetrically constrained transmitters because the condition $M_1, N_1 > N_2$ ensures that T1 can transmit a signal that violates this latter constraint (while R1 is still able to decode its desired signal). Thus, when $M_2 < N_2 < \min(M_1,N_1)$, one must consider a stricter constraint which, in essence, dictates that the interference at R2 can not be distributed arbitrarily in the receive signal-space of R2. This notion, which at $d_2 = M_2$ asserts that the interference is localized within some $(N_2-M_2)$-dimensional subspace, is referred to henceforth as {\em interference localization}. Indeed, it is because the outer-bounds derived in \cite{Chiachi2, D.Guo, Vaze_Dof_final} do not use this stronger constraint that they fail to characterize the DoF region of the ICs with asymmetrically constrained transmitters. Section \ref{subsec: intuition IC missing case} provides a more detailed heuristic explanation. On the other hand, in \cite{Zhu_Guo_noCSIT_DoF_2010}, the authors overcome this problem by first showing that it is DoF-region optimal for T2 to transmit $X^2(t)$ which is Gaussian with a covariance matrix that is proportional to the identity matrix. Consequently, with such an $X^2(t)$, they prove that the $M_2$-dimensional subspace spanned by $X^2(t)$ at R2 can not carry interference with a non-zero multiplexing gain. In a way, this latter point can be seen to implicitly capture the idea of interference localization described above.

In this paper, we provide a simpler and more generic proof of the result of \cite{Zhu_Guo_noCSIT_DoF_2010}. Unlike in \cite{Zhu_Guo_noCSIT_DoF_2010}, our proof does not require specialized techniques such as showing that the DoF-region optimality is retained by restricting $X_2$ to be Gaussian. Instead, the proof here makes use of basic information-theoretic identities such as the chain rules for differential entropy and mutual information, conditioning reduces entropy, etc. Consequently, the techniques developed here have the potential to be applicable for a wider class of networks.

As a case in point, we also study here the MIMO cognitive radio channel (CRC) \cite{Devroye}, which is defined as the MIMO IC with T2 cognitive (i.e., T2 knows the message of T1 as well). For the MIMO CRC, we determine the no-CSIT DoF region for the only class of MIMO CRCs for which the inner and outer-bounds established earlier by the authors in \cite{Vaze_Dof_final} were not tight, namely, that defined by the inequality $\min(M_1+M_2,N_1) > N_2 > M_2$. Our result here therefore completes the DoF region characterization of the MIMO CRC. In contrast, the applicability of the approach of \cite{Zhu_Guo_noCSIT_DoF_2010} is unclear, because it is not clear if the optimality of Gaussian $X_2$ can be proved in this problem, which is a critical step in the proof of \cite{Zhu_Guo_noCSIT_DoF_2010}. The DoF region of the MIMO CRC with CSIT was obtained in  \cite{Chiachi-Jafar}. The reader is also referred to \cite{Vaze_Dof_final} for a comparison of the DoF region with CSIT with the achievable DoF region of \cite{Vaze_Dof_final} which in turn we show to be the fundamental DoF region in this paper.

It is also shown in \cite{GDoF_region_noCSIT_IC_isit11} that the techniques of the present paper are also useful for characterizing the {\em generalized} degrees of freedom (GDoF) region \cite{Etkin_Tse_Wang} of the MIMO IC with asymmetrically constrained transmitters in the very weak interference regime.  Here again, it is unclear if the approach of \cite{Zhu_Guo_noCSIT_DoF_2010} is applicable.

The rest of the paper is organized as follows. Section \ref{sec: channel model missing case} presents the channel model and states the main results regarding the no-CSIT DoF regions of the IC and CRC with asymmetrically constrained transmitters (see Theorems \ref{thm: DoF region noCSIT IC missing case}-\ref{thm: DoF region noCSIT CRC correlated missing case}). Sections \ref{sec: proof of thm: DoF region noCSIT IC missing case}-\ref{sec: proof of thm: DoF region noCSIT CRC correlated missing case} present the proofs of those results with Section \ref{subsec: comparison with zhu guo} contrasting the proof technique developed here with that of \cite{Zhu_Guo_noCSIT_DoF_2010}. Section \ref{sec: conclusion missing case} concludes this paper.

\section{Channel Model, Definitions, and Main Results} \label{sec: channel model missing case}

The input-output relationship for the MIMO IC is given by equations (\ref{eq: IC model missing case R1}) and (\ref{eq: IC model missing case R2}). Note that the $(M_1,M_2,N_1,N_2)$ CRC is also governed by the same relationship, except that in the case of CRC, T2 is cognitive in the sense that it knows the message of T1. We now state our assumptions about the distributions of the additive noises and channel matrices.

We let the elements of the additive noises $W(t)$ and $W'(t)$ be independent and identically distributed (i.i.d.) according the circularly symmetric complex Gaussian distribution with zero mean and unit variance, denoted henceforth as $\mathcal{C}\mathcal{N}(0,1)$. The noise as well as the channel realizations are assumed to be i.i.d. across time. Moreover, all channel matrices and additive noises are taken to be independent.

Further, we assume that both the receivers know all channel matrices perfectly and instantaneously but the transmitters know only their distribution. This assumption is referred as the `no CSIT' assumption.

We introduce some notation. Let $\mathbf{H}^b \define \big\{ H^{11}(t), H^{12}(t), H^{21}(t), H^{22}(t) \big\}_{t=1}^b$, $\mathbf{Y}^b \define \{Y(t)\}_{t=1}^b$, and $\mathbf{Z}^b \define \{Z(t)\}_{t=1}^b$. Further, define a binary-valued variable $1_{T2}$ which takes value $1$ if T2 is cognitive, else it is zero. In other words, $1_{T2} = 1$ only when we are dealing with the CRC. For any random variable $V$, we define $1_{T2} V = V$ if $1_{T2} = 1$, else $1_{T2} V = 0$.

Let $M_Y$ and $M_Z$ be two independent messages, which are intended for R1 and R2, respectively, and are to be sent by the transmitters over a block of length $b$. It is assumed that $\mathcal{M}_i$ is distributed uniformly over a set of cardinality $2^{nR_i(P)}$, when there is a power constraint of $P$ at the transmitters. A coding scheme for blocklength $b$ consists of two encoding functions $f^{(i,b)} = \{f_t^{(i,b)}\}_{t=1}^b$, $i = 1,2$, given as
\begin{eqnarray*}
X^1(t) & = & f_t^{(1,b)} \big( M_Y \big) \; {\rm and } \\
X^2(t) & = & f_t^{(2,b)} \big( M_Z, 1_{T2} M_Y \big) ,
\end{eqnarray*}
$  \forall t \in \{1,2,\cdots,b\} $ and two decoding functions defined as
\begin{eqnarray*}
\hat{M}_Y & = & g^{(1,b)} \big( \mathbf{Y}^b, \mathbf{H}^b \big) \; {\rm and} \\
\hat{M}_Z & = & g^{(2,b)} \big( \mathbf{Z}^b, \mathbf{H}^b \big).
\end{eqnarray*}
A rate tuple $\big( R_1(P),R_2(P) \big)$ is said to be achievable if there exists a sequence of coding schemes, one for each $b$, such that the probability of $M_Y \not= \hat{M}_Y$ or $M_Z \not= \hat{M}_Z$ tends to zero as $b \to \infty$.

The capacity region $\mathcal{C}(P)$ is defined as the set of all rate tuples that are achievable when there is a power constraint of $P$ at T1 and T2. If $\mg (x) \define \lim_{P \to \infty} \frac{x}{\log_2 P},$ then the DoF region is defined for now as
\begin{eqnarray*}
\label{eq:defn-dof}
\lefteqn{ \hspace{-0.9cm}  \mathbf{D}_{\rm{IC \backslash CRC}} = \Big\{ (d_1,d_2) \Big|  \exists \mbox{ a sequence } \Big< \big( R_1(P),R_2(P) \big) \Big>_P  \Big. \Big. } \\
&& {} \Big. \in \mathcal{C}(P) \mbox{such that } 0 \leq d_i = \mg \big( R_i(P) \big) ~ \forall i \Big\}.
\end{eqnarray*}

Note that the above definition of the DoF region is restrictive in the sense that a DoF pair $(d_1,d_2) \in \mathbf{D}_{\rm{IC \backslash CRC}}$ only if $d_i$ is the limit of the sequence $\left< \frac{R_i(P)}{\log_2 P} \right>_P$, $i = 1,2$. The existence of these limits however puts an implicit but undue constraint on the inputs. In Section \ref{subsec: DoF region more general missing case}, we define the DoF region more generally using the limit superior \cite{Royden} (cf. \cite{JaferShamai}) and prove that this constraint does not result in a larger ``true" DoF region. Until then, the use of the definition in \eqref{eq:defn-dof} allows us to keep the explanation of the key ideas of the proof relatively simple.

\subsection{Some Definitions}
To specify the distributions of the channel matrices, we make use of the following definitions.
\begin{definition}[\cite{Zhu_Guo_noCSIT_DoF_2010}]
An $M \times N$ random matrix $H$ is said to be isotropic if $H$ and $HU$ have the same distribution (denoted symbolically as $H \sim HU$) for any deterministic $N \times N$ unitary matrix $U$.
\end{definition}

\begin{definition}[isotropic fading]
The channel matrices are said to be isotropically distributed if all channel matrices are isotropically distributed, i.e., $H^{ij}(t)$ is isotropically distributed for all $t$, $i$, and $j$.
\end{definition}

\begin{definition}[i.i.d. Rayleigh fading]
\label{def:iid}
The channel matrices are said to be i.i.d. Rayleigh-faded if all entries of all channel matrices $\{H^{ij}(t)\}_{i,j}$ are i.i.d. (across $i$, $j$, and $t$) according to $\mathcal{C}\mathcal{N}(0,1)$ distribution.
\end{definition}
Note that if the channel matrices are i.i.d. Rayleigh-faded then they are also isotropically distributed, but not necessarily otherwise.

We now define a specific type of correlated Rayleigh fading. Let the entries of $H_w^{11}(t)$ and $H_w^{21}(t)$ be i.i.d. $\mathcal{C}\mathcal{N}(0,1)$ random variables. Further, consider two matrices $H^{12}_w(t)$ and $H^{22}_w(t)$ of sizes $N_1 \times M_2$ and $N_2 \times M_2$, respectively, such that the first $N_1-(N_2-M_2)$ and $N_2-(N_2-M_2)$ rows of them (resp.) consist of i.i.d. $\mathcal{C}\mathcal{N}(0,1)$ random variables, with the last $(N_2-M_2)$ rows consisting only of zeros.

\begin{definition}[correlated Rayleigh fading]
\label{def:cor}
The channel matrices are said to follow correlated Rayleigh fading if, for each $i \in \{1,2\}$, $H^{i1}(t) \sim H^{i1}_w(t)$ and $H^{i2}(t) \sim U^{i2} H^{i2}_w(t)$ for some deterministic $N_i \times N_i$ unitary matrix $U^{i2}$.
\end{definition}
Note that the channel matrices are full rank under correlated Rayleigh fading.

The following definition helps us state the DoF regions of the IC and the CRC.

\begin{definition}
For an integer-valued function $N_1'$ of $(M_1,M_2,N_1,N_2)$,
\begin{eqnarray*}
\lefteqn{ \mathbf{D}(M_1,M_2,N_1,N_2,N_1') \define \Big\{ (d_1,d_2) \Big| L_{o1} \equiv 0 \leq d_1 \leq N_1', \Big. \Big. } \\
&& {} \hspace{-10pt} \Big. L_{o2} \equiv 0 \leq d_2 \leq M_2, ~ L \equiv d_1 + \frac{N_1'+M_2 - N_2}{M_2} d_2 \leq N_1' \Big\}.
\end{eqnarray*}
\end{definition}
The three bounds appearing in the above definition are henceforth referred to as $L_{o1}$, $L_{o2}$, and $L$, respectively.

\subsection{Main Results}

The following theorem states the no-CSIT DoF region of the IC with asymmetrically constrained transmitters under isotropic fading.
\begin{theorem} \label{thm: DoF region noCSIT IC missing case}
For the MIMO IC with isotropic fading and $(M_1,M_2,N_1,N_2)$ such that the inequality $\min(M_1,N_1) > N_2 > M_2$ holds, the no-CSIT DoF region, $\mathbf{D}_{\rm{IC}}$, is equal to the region $\mathbf{D}(M_1,M_2,N_1,N_2,N_1')$ with $N_1' = \min(N_1,M_1)$, i.e.,
\[
\mathbf{D}_{\rm{IC}} = \mathbf{D} \big( M_1,M_2,N_1,N_2,\min\{N_1,M_1\} \big).
\]
\end{theorem}
\begin{IEEEproof}
Bound $L$ intersects bounds $L_{o1}$ and $L_{o2}$ at points $(N_1',0)$ and $(N_2-M_2,M_2)$, respectively. These points are achievable via simple receive zero-forcing (cf. \cite[Theorem 4]{Vaze_Dof_final}). Hence, the region $\mathbf{D}(M_1,M_2,N_1,N_2,N_1')$ with $N_1' = \min(N_1,M_1)$ is achievable via receive zero-forcing and time sharing. On the converse side, $L_{o1}$ and $L_{o2}$ are outer-bounds since the number of DoF achievable over the point-to-point MIMO channel can not exceed the minimum of the number of transmit and receive antennas (henceforth called the single-user bound) \cite{Telatar}. It is thus sufficient to establish that $L$ is an outer-bound. The detailed proof of this claim, which is different and simpler than the one given by \cite{Zhu_Guo_noCSIT_DoF_2010}, is given in Section \ref{sec: proof of thm: DoF region noCSIT IC missing case}.
\end{IEEEproof}

We next consider the MIMO CRC with $\min(M_1+M_2,N_1) > N_2 > M_2$, which is henceforth referred to as the CRC with asymmetrically constrained transmitters. Its DoF region is determined below for the cases of the i.i.d. Rayleigh fading and correlated Rayleigh fading models.

\begin{theorem} \label{thm: DoF region noCSIT CRC iid missing case}
For the MIMO CRC with i.i.d. Rayleigh fading of Definition \ref{def:iid} and $(M_1,M_2,N_1,N_2)$ such that the inequality $\min(M_1+M_2,N_1) > N_2 > M_2$ holds, the no-CSIT DoF region, $\mathbf{D}_{\rm{CRC}}$, is equal to the region $\mathbf{D}(M_1,M_2,N_1,N_2,N_1')$ with $N_1' = \min(N_1,M_1+M_2)$, i.e.,
\[
\mathbf{D}_{\rm{CRC}} = \mathbf{D} \big( M_1,M_2,N_1,N_2,\min\{N_1,M_1+M_2\} \big).
\]
\end{theorem}
\begin{IEEEproof}
Achievability follows by noting that the bound $L$ in this case passes through points $(N_1',0)$ and $(N_2-M_2,M_2)$, both of which can be achieved by simple receive zero-forcing (cf. \cite[Theorem 7]{Vaze_Dof_final}). On the converse side, as argued before, it is sufficient to prove that $L$ is an outer-bound, which is done in Section \ref{sec: proof of thm: DoF region noCSIT CRC iid missing case}.
\end{IEEEproof}

Using the above theorem and the results of \cite{Vaze_Dof_final}, we can now state the DoF region of the CRC with i.i.d. Rayleigh fading.
\begin{theorem}
The DoF region of the MIMO CRC with i.i.d. Rayleigh fading and no CSIT is given by
\begin{eqnarray*}
\lefteqn{ \mathbf{D}_{\rm{CRC}} = \Big\{ (d_1, d_1) ~ \big| ~ 0 \leq d_1 \leq \min \{ N_1,M_1 + M_2; \}; \Big. } \\
&& {} 0 \leq d_2 \leq \min \{M_2,N_2\}; \mbox{ if } N_1 \leq N_2 \mbox{ then} \\
&& {} \frac{d_1}{\min(N_1,M_2)} + \frac{d_2}{\min(N_2,M_2)} \leq \frac{\min(N_1,M_1 + M_2)}{\min(N_1,M_2)}; \\
&& {} \hspace{-5pt} \mbox{ if } \min(M_1+M_2,N_1) > N_2 > M_2 \mbox{ then} \\
&& {} d_1 + d_2 \frac{N_1' + M_2 - N_2}{M_2} \leq N_1' , \\
&& {} \Big. \mbox{else } \frac{d_1}{\min(N_1,M_1+M_2)} + \frac{d_2}{\min(N_2,M_1+M_2)} \leq 1 \Big\},
\end{eqnarray*}
where $N_1' = \min(N_1,M_1+M_2)$.
\end{theorem}
\begin{IEEEproof}
Follows from Theorem \ref{thm: DoF region noCSIT CRC iid missing case} above, and Theorems 7 and 8, and Remark 21 of \cite{Vaze_Dof_final}.
\end{IEEEproof}

\begin{theorem} \label{thm: DoF region noCSIT CRC correlated missing case}
For the MIMO CRC with correlated Rayleigh fading of Definition \ref{def:cor} and $(M_1,M_2,N_1,N_2)$ such that the inequality $\min(M_1+M_2,N_1) > N_2 > M_2$ holds, the no-CSIT DoF region, $\mathbf{D}_{\rm{CRC}}$, is equal to the region $\mathbf{D}(M_1,M_2,N_1,N_2,N_1')$ with $N_1' = \min(N_1,M_1+M_2)$, i.e.,
\[
\mathbf{D}_{\rm{CRC}} = \mathbf{D} \big( M_1,M_2,N_1,N_2,\min\{N_1,M_1+M_2\} \big).
\]
\end{theorem}
\begin{IEEEproof}
Again the achievability follows from simple receive zero-forcing (cf. the proof of Theorem \ref{thm: DoF region noCSIT CRC iid missing case}). On the converse side, again, it is sufficient to prove that $L$ is an outer-bound, which is shown in Section \ref{sec: proof of thm: DoF region noCSIT CRC correlated missing case}.
\end{IEEEproof}

\section{Proof of Theorem \ref{thm: DoF region noCSIT IC missing case}: $L$ is an Outer-Bound} \label{sec: proof of thm: DoF region noCSIT IC missing case}

Before starting the proof, we introduce some notation. \newline
\emph{\underline{Notation:} } For a column vector $V(t)$ define
$\mathbf{V} \equiv \mathbf{V}_1^b$ to be a vector $[V^T(1) V^T(2) \cdots V^T(b)]^T $, where $[\cdot]^T$ denotes a transpose of a matrix or vector. Let $V_i(t)$ denote the $i^{th}$ element of the column vector $V(t)$. Similarly, for a matrix $H^{kl}(t)$, $H^{kl}_i(t)$ denotes its $i^{th}$ row. Define $(\mathbf{V}_i)_1^b \equiv \mathbf{V}_i$ to be the vector $ [ V_i(1), V_i(2), \cdots , V_i(b) ]^T$. Further, for integers $n_1$ and $n_2$ with $n_1 \leq n_2$, let $[n_1:n_2] \define \{n_1, n_1+1,\cdots, n_2\}$, $\mathbf{V}_{[n_1:n_2]}\define \{ \mathbf{V}_{j}\}_{j=n_1}^{n_2}$, and
\[
H^{ij}_{[n_1:n_2]}(t) = \begin{bmatrix} H^{ij}_{n_1}(t) \\ H^{ij}_{n_1+1}(t) \\ \vdots \\ H^{ij}_{n_2}(t) \end{bmatrix}.
\]
Following \cite{Royden}, for a real-valued sequence $\big< \!  x_n  \! \big>_n$, limit superior, limsup, is defined as
\[
\overline{\lim_{n \to \infty}} ~ x_n = \inf_n \sup_{n_0 \geq n} x_{n_0} = \lim_{n \to \infty} \sup_{n_0 \geq n} x_{n_0}.
\]
Then, for a real-valued function $x(b,P)$ of $b$ and $P$, let
\[
f(x) \define \overline{\lim_{P \to \infty}} ~  \frac{1}{\log_2 P} \Big\{ \overline{\lim_{b \to \infty}} ~ \frac{x}{b} \Big\}.
\]
Note that the function $f(\cdot)$ preserves the sense of inequality. Finally, $\mathbf{H}_1^b \equiv \mathbf{H} \define \big\{ \{H^{ij}(t)\}_{i,j=1}^2 \big\}_{t=1}^b$. \vspace{1mm}

In Section \ref{subsec: intuition IC missing case}, the intuition behind the proof is explained. Using this insight, the main result is proved in Section \ref{subsec: main proof IC missing case}.

\subsection{Interference localization: An intuitive explanation} \label{subsec: intuition IC missing case}
As stated earlier, \cite{Chiachi2, D.Guo, Vaze_Dof_final} provided an (identical) outer-bound to the no-CSIT DoF region of the MIMO IC. However, that outer-bound turns out to be loose for the ICs with asymmetrically constrained transmitters. In what follows, we briefly explain the technique of \cite{Chiachi2, D.Guo, Vaze_Dof_final} that results in the (common) outer-bound and then describe why this bound fails to yield the exact DoF region for this class of ICs. Following that, we outline how the tight outer-bound of this paper is derived.

In \cite{Chiachi2, D.Guo, Vaze_Dof_final}, the outer-bound is derived by applying Fano's inequality at R2, which, after some manipulations, yields the implication
\begin{equation}
d_2 \mbox{ is achievable } \Rightarrow f \big( ~ I(\mathbf{X^1} ; \mathbf{Z} | M_Z, \mathbf{H}) ~ \big) \leq N_2 - d_2 \label{eq: bound on Interf. at R2 preliminary CRC intuition missing case}
\end{equation}
(see the derivation of (\ref{eq: bound on Interf. at R2 preliminary CRC missing case}) in the next sub-section). The inequality in \eqref{eq: bound on Interf. at R2 preliminary CRC intuition missing case} puts constraints on the transmission scheme of T1. Using this fact, \cite{Chiachi2, D.Guo, Vaze_Dof_final} upper-bound the achievable value of $d_1$ in terms of a function of $d_2$ and $(M_1,M_2,N_1,N_2)$, from which the outer-bound is computed therein.

Note that $I(\mathbf{X^1}; \mathbf{Z} | M_Z, \mathbf{H})$ is a measure of the interference seen by R2, and the inequality in (\ref{eq: bound on Interf. at R2 preliminary CRC intuition missing case}) upper-bounds the multiplexing gain of the total interference seen by R2 per unit time. Therefore, the outer-bound of \cite{Chiachi2, D.Guo, Vaze_Dof_final}, which is based on (\ref{eq: bound on Interf. at R2 preliminary CRC intuition missing case}), is referred to henceforth as the {\em total interference} outer-bound.

It turns out that although the implication in (\ref{eq: bound on Interf. at R2 preliminary CRC intuition missing case}) holds, its reverse implication may not for MIMO ICs with \act. More precisely, for this class of ICs, it is possible that
\begin{equation}
f \big( ~ I(\mathbf{X^1} ; \mathbf{Z} | M_Z, \mathbf{H}) ~ \big) \leq N_2 - d_2 \not\Rightarrow d_2 \mbox{ is achievable}. \label{eq: bound on Interf. at R2 preliminary CRC intuition insufficiency of the bound missing case}
\end{equation}
Thus, the total interference outer-bound 
fails to characterize the DoF region.

\begin{figure} \centering
\includegraphics[bb=180bp 55bp 520bp 500bp,clip,scale=0.4]{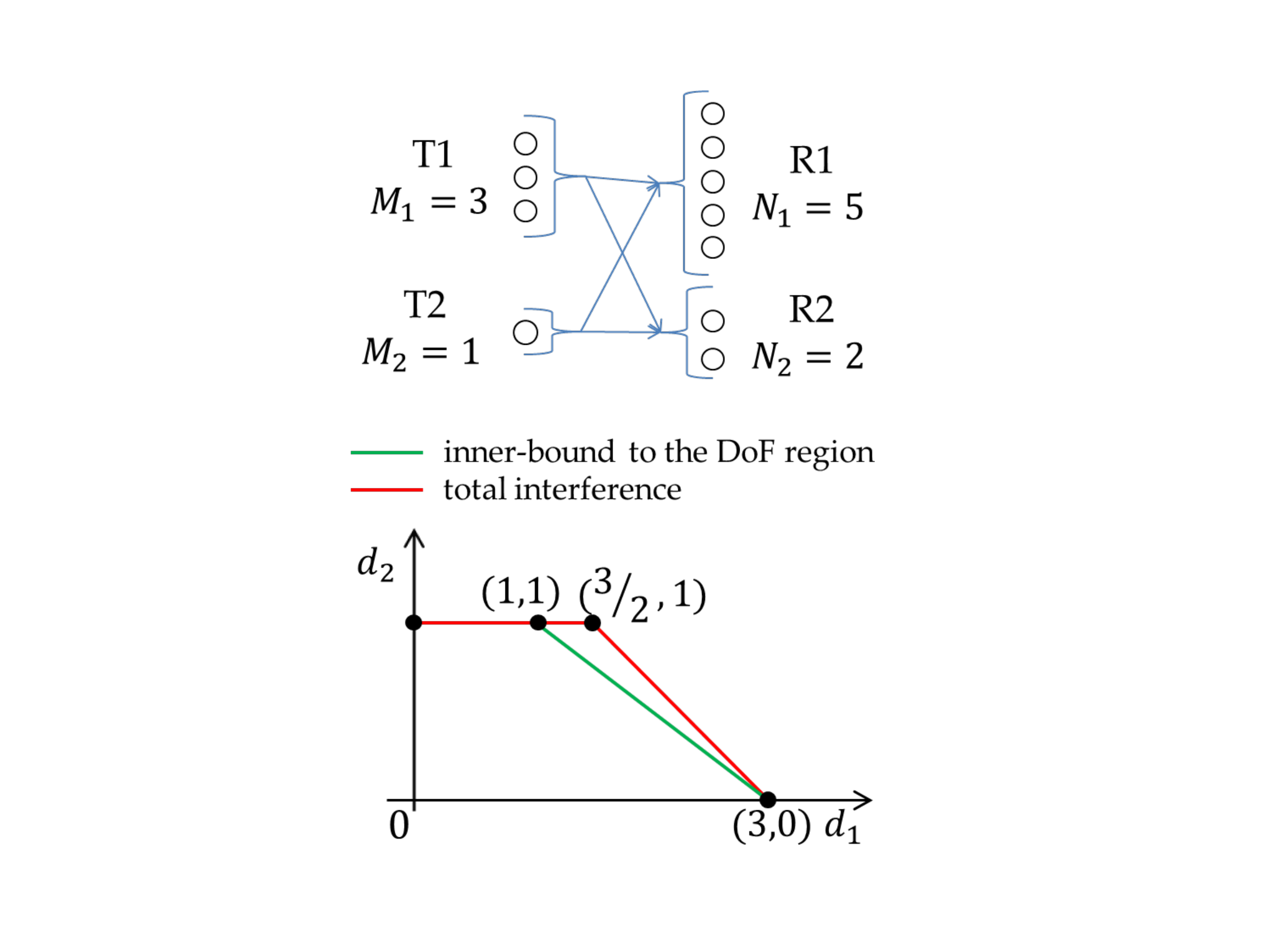}
\caption{An Example of the IC with Asymmetrically Constrained Transmitters} \label{fig: example_IC_intuition_IC_missingcase}
\end{figure}

\begin{figure} \centering
\includegraphics[bb=150bp 110bp 650bp 410bp,clip,scale=0.4]{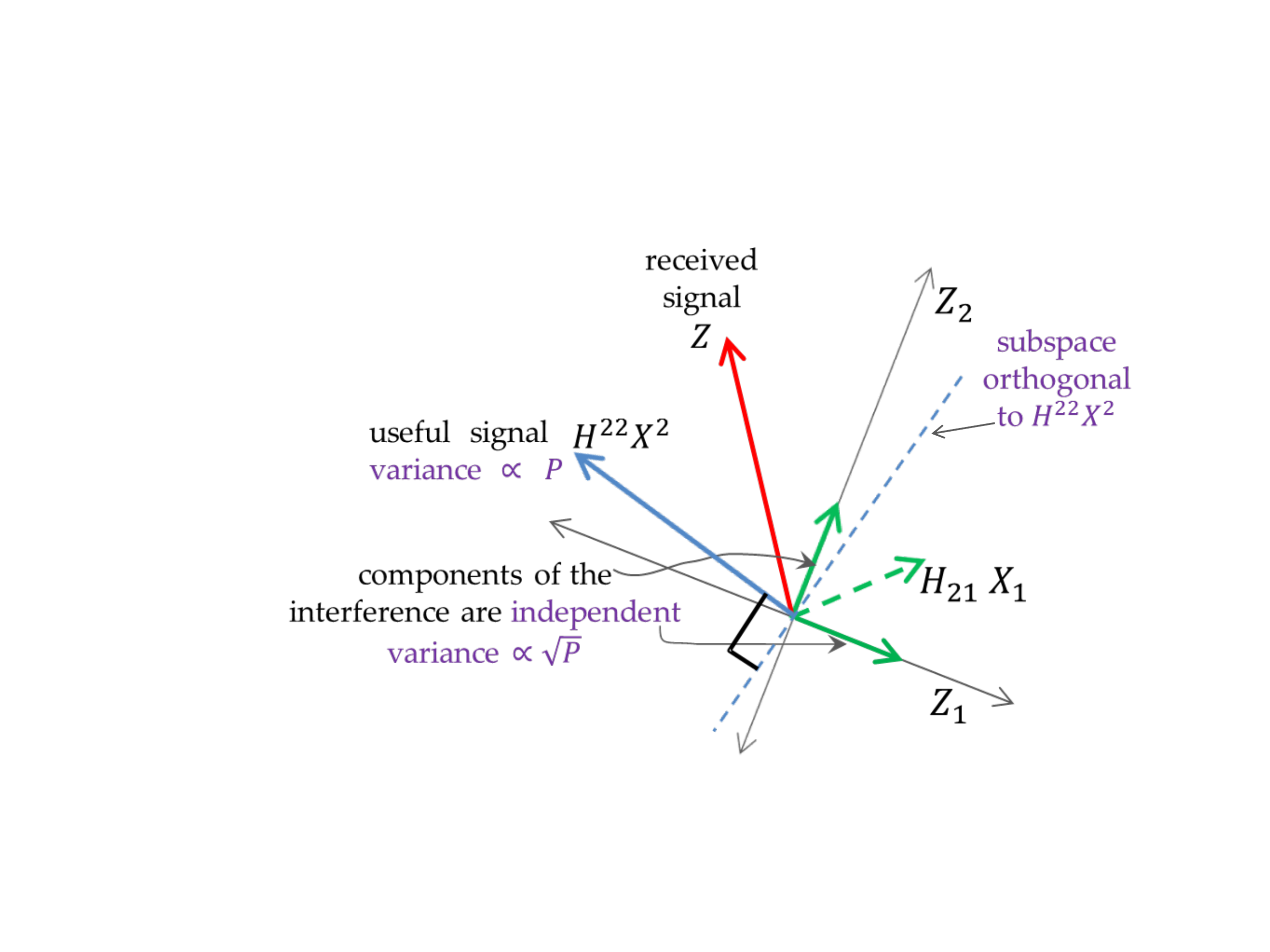}
\caption{The Receive Signal-Space of R2 with Uniform Signaling at T1}
\label{fig: receive signal space of R2 intuition-example missing case}
\end{figure}

To understand this, let us consider an example of the IC with $(M_1,M_2,N_1,N_2) = (3,1,5,2)$ (see Fig. \ref{fig: example_IC_intuition_IC_missingcase}) and focus on the case of $d_2 = 1$. The inequality in (\ref{eq: bound on Interf. at R2 preliminary CRC intuition missing case}) reduces to
\begin{equation}
f \big( ~ I(\mathbf{X^1} ; \mathbf{Z} | M_Z, \mathbf{H}) ~ \big) \leq 1. \label{eq: bound on Interf. at R2 preliminary CRC intuition-example missing case}
\end{equation}
Consider a particular transmission scheme which satisfies the above inequality. Suppose T1 transmits $3$ data symbols that are i.i.d. according to $\mathcal{C}\mathcal{N}(0,P^{\alpha})$, $0 <\alpha \leq \frac{1}{2}$ (we refer to such signaling as {\em uniform signaling});  T2 transmits a $\mathcal{C}\mathcal{N}(0,P)$ data symbol; and the signals of T1 and T2 are i.i.d. across time. It is not difficult to prove that such a strategy satisfies the inequality in (\ref{eq: bound on Interf. at R2 preliminary CRC intuition-example missing case}) (with equality if $\alpha = \frac{1}{2}$). Moreover, R1 has sufficient number of antennas to zero-force the interference and achieve up to $3 \alpha \leq \frac{3}{2}$ DoF. Now consider the receive signal-space of R2 shown in Fig. \ref{fig: receive signal space of R2 intuition-example missing case} where, for simplicity, we take $b = 1$ and time index $t$ is shown explicitly. With uniform signaling at T1, the interference $H^{21}(t) X^1(t)$ at R2 satisfies the following properties\footnote{These properties can be easily proved for $H^{21}(t) = \begin{bmatrix} I_2 & 0_{2 \times 1} \end{bmatrix}$. In the general case, R2 can apply an invertible transformation on the received signal $Z(t)$ to compute $\underline{Z}(t) = \underline{H^{22}}(t) X^2(t) + \underline{H^{21}}(t) X^1(t) + \underline{W'}(t)$, where  $\underline{H^{21}}(t) = \begin{bmatrix} I_2 & 0_{2 \times 1} \end{bmatrix}$. Since $\underline{Z}(t)$ is obtained from $Z(t)$ using an invertible transformation, the mutual information terms would remain unchanged, i.e., $I(\mathbf{X^1} ; \mathbf{Z} | M_Z, \mathbf{H}) = I(\mathbf{X^1} ; \mathbf{\underline{Z}} | M_Z, \mathbf{H})$. Therefore, we can regard $\underline{Z}(t)$ as the signal received at R2, and the stated properties can be proved for this equivalent channel.}:
\begin{enumerate}[(a)]
\item If we pick any $2$ orthonormal basis vectors for the $2$-dimensional receive signal space of R2, then the components of the interference along the two basis vectors are independent and each has a variance of $P^{\alpha}$; and
\item any $1$-dimensional subspace chosen in the $2$-dimensional receive signal space of R2 carries a component of the interference with multiplexing gain equal to $\alpha$ (since its variance is $P^{\alpha}$).
\end{enumerate}
Further, the useful signal $H^{22}(t) X^2(t)$ can span only a $1$-dimensional subspace since $M_2 =1$. Hence, the subspace orthogonal to the span of $H^{22}(t)$ can not give any information to R2 about the useful signal $X^2(t)$. In other words, out of the total $2$ DoF available to R2, $1$ DoF is lost because T2 has just $M_2 = 1 (< 2 = N_2)$ antenna. Moreover, out of the $1$ DoF that is left at R2, $\alpha$ DoF are occupied by the interference (see Property (b) of the interference at R2). Thus, R2 has only $1 - \alpha$ DoF available for decoding the useful signal so that it can not achieve $d_2 = 1$. Hence the claim of (\ref{eq: bound on Interf. at R2 preliminary CRC intuition insufficiency of the bound missing case}) is true.

We next argue through the same example that the claim of (\ref{eq: bound on Interf. at R2 preliminary CRC intuition insufficiency of the bound missing case}) holds because $M_2 < N_2$. Now, R2 can not achieve $d_2 = 1$ since the $1$ DoF available to it is lost due to the limitation at T2 that its transmit signal can not span the entire receive signal space. In particular, had this limitation not existed such as when $M_2=2$ (with $M_1$, $N_1$, and $N_2$ unchanged), R2 would have been able to achieve $d_2 = 1$. To see this, note that T2 in this case can transmit two complex Gaussian symbols each with a power of $\frac{P}{2}$ and make $X^2(t)$ span the entire receive signal space enabling R2 to achieve $d_2 = 1$ by treating interference as noise, even if T1 employs uniform signalling (note R1 can still zero-force the interference to successfully recover the useful signal). Therefore, we conclude that the claim in (\ref{eq: bound on Interf. at R2 preliminary CRC intuition insufficiency of the bound missing case}) holds because $M_2 < N_2$. Hence the implication in (\ref{eq: bound on Interf. at R2 preliminary CRC intuition missing case}) is insufficient in the sense that it does not capture the further limitation imposed by $M_2 < N_2$.

\begin{figure} \centering
\includegraphics[bb=150bp 140bp 610bp 380bp,clip,scale=0.4]{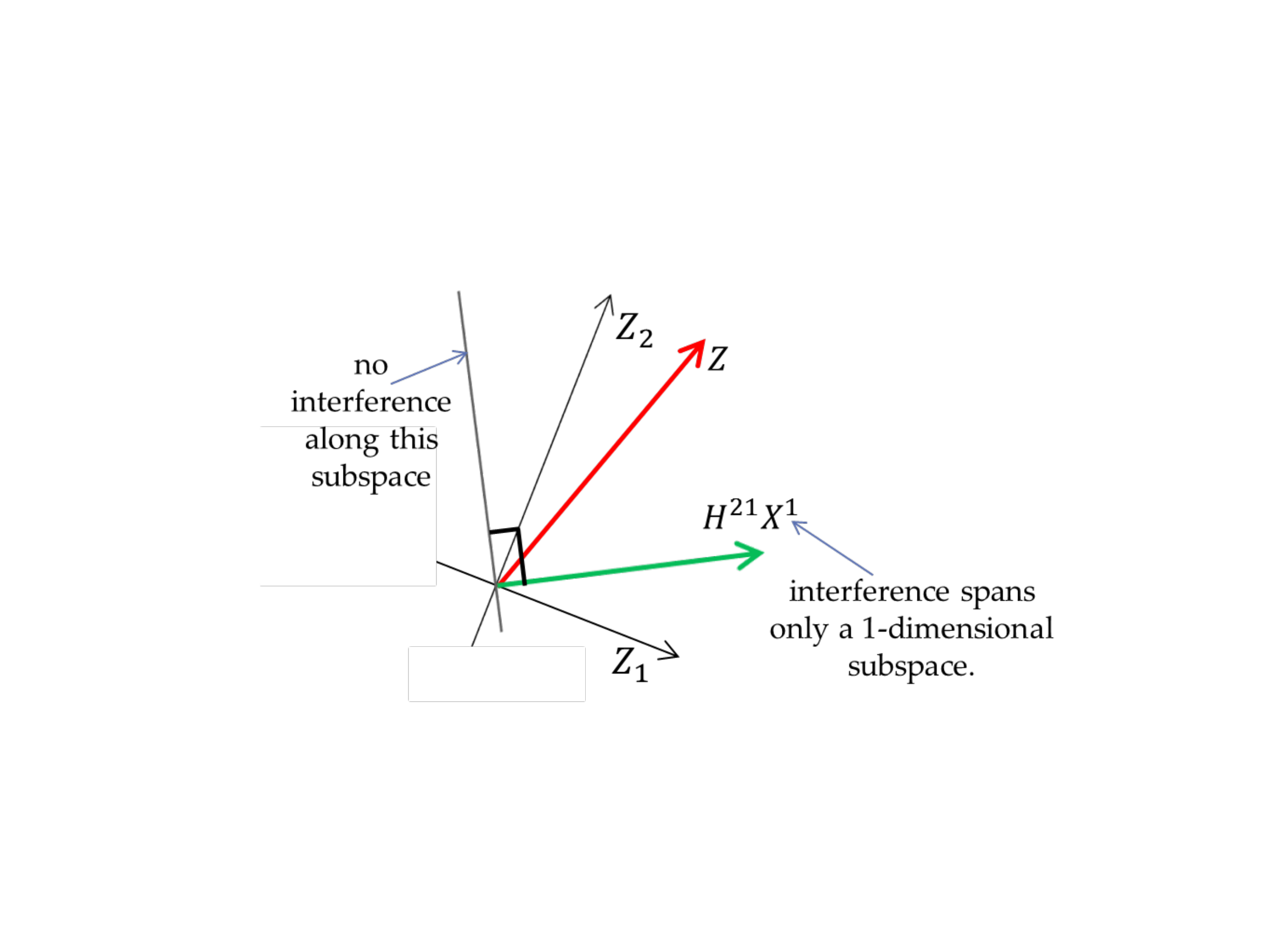}
\caption{Interference Localization at R2}
\label{fig: interference localizn at R2 intuition-example missing case}
\end{figure}

Indeed, for ICs with \act, we must constrain how the interference is distributed in the receive signal-space of R2 in addition to upper-bounding its multiplexing gain using the inequality of \eqref{eq: bound on Interf. at R2 preliminary CRC intuition insufficiency of the bound missing case}. This is explained in the context of our example. It must be proved that if $d_2 = 1$ is achievable then the interference $H^{21}(t) X^1(t)$ spans a $1$-dimensional subspace at R2, or there exists a subspace which does not contain any interference (with positive multiplexing gain), see Fig. \ref{fig: interference localizn at R2 intuition-example missing case}. This is because if this were not true, then, as argued for the case where T1 employs uniform signaling, $d_2 = 1$ can not be achieved. In other words, we must prove that if $d_2 =1$ is achievable, then the interference is localized to a smaller-dimensional subspace and it {\em cannot} be distributed uniformly in the receive signal-space of R2, which is the case if T1 employs uniform signaling.

In general, it must be shown that if $d_2$ DoF are achievable for T2-R2 pair over an IC with \act, then the interference $H^{21}(t) X^1(t)$ at R2 must be such that
\begin{enumerate}[(a)]
\item its multiplexing gain is at most $N_2 - d_2$ (as required by the inequality in (\ref{eq: bound on Interf. at R2 preliminary CRC intuition insufficiency of the bound missing case})); and {\em additionally},
\item there exists an $M_2$-dimensional subspace in the receive signal-space of R2 that carries interference with multiplexing gain at most $M_2 - d_2$.
\end{enumerate}
We call this property {\em interference localization}, because at $d_2 = M_2$, it amounts to the entire interference being localized to some $(N_2 - M_2)$-dimensional subspace. Our intuition suggests that if this property is proved, we would get the tightest characterization of the DoF region. Indeed, Lemma \ref{lem: inter. localizn IC missing case} of the next sub-section accomplishes this task, using which the desired bound $L$ is derived.

\subsection{Main Proof} \label{subsec: main proof IC missing case}
We prove here that for the MIMO IC with \act, $L$ with $N_1' = \min (M_1,N_1)$ is an outer-bound. To this end, first obtain the singular-value decomposition of the isotropically-distributed channel matrices.
\begin{lemma}
For an $N_i \times M_j$ isotropically-distributed channel matrix $H^{ij}(t)$, we may write
\[
H^{ij}(t) = U^{ij}(t) \Lambda^{ij}(t) \big( V^{ij}(t) \big)^*,
\]
where $U^{ij}(t)$, $\Lambda^{ij}(t)$, and $V^{ij}(t)$ are deterministic functions of $H^{ij}(t)$ such that
\begin{enumerate}[(i)]
\item $U^{ij}(t)$ is an $N_i \times N_i$ unitary matrix;
\item $\Lambda^{ij}(t)$ is an $N_i \times \min(N_i,M_j)$ diagonal matrix containing the singular values of $H^{ij}(t)$, i.e., the square matrix formed by retaining its first $\min(N_i,M_j)$ rows, denoted henceforth as $\tilde{\Lambda}^{ij}(t)$, is diagonal with singular values of $H^{ij}(t)$ along its diagonal and the remaining rows consist only of zeros;
\item $V^{ij}(t)$ is an $M_j \times \min(N_i,M_j)$ isotropically-distributed semi-unitary matrix, i.e., $\big(V^{ij}(t)\big)^* V^{ij}(t) = I_{\min(N_i,M_j)}$ and it is uniformly distributed over its domain; and
\item $V^{ij}(t)$ is independent of $U^{ij}(t)$ and $\Lambda^{ij}(t)$.
\end{enumerate}
\end{lemma}
\begin{IEEEproof}
Follows from \cite[Lemma 1]{Zhu_Guo_noCSIT_DoF_2010}.
\end{IEEEproof}

To explain the main idea, we first consider the case where $U^{11}(t) = I_{N_1}$, $U^{21}(t) = I_{N_2}(t)$, and all singular values of $H^{11}(t)$ and $H^{21}(t)$ are equal to unity with probability $1$. These assumptions about $U^{i1}(t)$ and $\Lambda^{i1}(t)$ will be in effect until the general case is discussed towards the end. Note that under these assumptions $\Lambda^{21}(t) = I_{N_2}$ since $N_2 < M_1$.

The proof now consists of three steps. \newline $\bullet$ Step I: Use Fano's inequality to bound $d_2$. It is argued that this bound can not immediately be used to obtain the desired bound $L$, which motivates the analysis of the next step. \newline $\bullet$ Step II: Obtain tight bounds on the interference at R2 by proving interference localization. \newline $\bullet$ Step III: Apply Fano's inequality to bound $d_1$ in terms of the multiplexing gain of a certain mutual information term (see equation (\ref{eq: basic bound on d1 missing case})), which is then upper-bounded using the bounds derived at Step II.

\underline{Step I:} We apply Fano's inequality \cite{CT} at R2 to obtain
\begin{equation}
b R_2  \leq  I(M_Z ; \mathbf{Z} | \mathbf{H}) + b \epsilon_b, \label{eq: bound on R2 missing case}
\end{equation}
where $b$ is the bocklength and $\epsilon_b \to 0$ as $b \to \infty$. This yields
\begin{eqnarray}
b R_2 & \hspace{-7pt}  \leq \hspace{-7pt} & I(M_Y , M_Z ; \mathbf{Z} | \mathbf{H}) - I(M_Y ; \mathbf{Z} | M_Z, \mathbf{H}) + b \epsilon_b \nonumber \label{eq: bound on R2 missing case basic form} \\
\Rightarrow b R_2 & \hspace{-7pt} + \hspace{-7pt} & I(M_Y ; \mathbf{Z} | M_Z, \mathbf{H}) \leq  I(M_Y , M_Z ; \mathbf{Z} | \mathbf{H}) + b \epsilon_b. \label{eq: bound on R2 missing case limsup form}
\end{eqnarray}

Now, if $(d_1,d_2) \in \mathbf{D}_{\rm{IC}}$, then, by definition, there exists a sequence $\big( R_1(P),R_2(P) \big) \in \mathcal{C}(P)$ such that $d_2 = \lim_{P \to \infty} \frac{1}{\log_2 P} \lim_{b \to \infty} \frac{b R_2}{b}$. Moreover, for any rate pair $\big( R_1(P),R_2(P) \big) \in \mathcal{C}(P)$, $R_2(P) \equiv R_2$ satisfies inequality (\ref{eq: bound on R2 missing case limsup form}). Therefore, from (\ref{eq: bound on R2 missing case limsup form}), we get
\begin{eqnarray}
d_2 + f \big( ~ I(M_Y ; \mathbf{Z} | M_Z, \mathbf{H}) ~ \big) & \hspace{-7pt} \leq \hspace{-7pt} & f \Big( b R_2 + I(M_Y ; \mathbf{Z} | M_Z, \mathbf{H}) \Big)  \nonumber \\
& \hspace{-7pt} \leq \hspace{-7pt} & f \Big(  I(M_Y , M_Z ; \mathbf{Z} | \mathbf{H}) + b \epsilon_b \Big) \nonumber \\
& \hspace{-7pt} =  \hspace{-7pt} & f \Big(  I(M_Y , M_Z ; \mathbf{Z} | \mathbf{H}) \Big) \nonumber \\
\Rightarrow d_2 \leq N_2 & \hspace{-7pt} - \hspace{-7pt} &  f \big( ~ I(M_Y ; \mathbf{Z} | M_Z, \mathbf{H}) ~ \big), \label{eq: bound on Interf. at R2 preliminary CRC missing case}
\end{eqnarray}
where the last inequality holds due to the single-user bound. Here, the number $f \big( I(M_Y ; \mathbf{Z} | M_Z, \mathbf{H}) \big)$ is equal to the multiplexing gain of the net (per unit time) interference encountered by R2; and the above inequality constrains the multiplexing gain of the total interference seen by R2 per unit time. However, as explained in the last sub-section, this inequality does not completely capture the limitation of the second transmit-receive pair due to $M_2 < N_2$. As per the discussion therein, we must prove an additional bound that constrains how the interference is distributed. Such bounds are derived in the following lemma.
%
%
%

\underline{Step II:} 
\begin{lemma}[Interference Localization] \label{lem: inter. localizn IC missing case}
We have
\begin{eqnarray}
f \big( ~ I(M_Y; ~ \mathbf{Z}_{[1:M_2]} ~  \big| M_Z, ~ \mathbf{Z}_{[M_2+1:N_2]}, ~ \mathbf{H}) ~ \big) \leq M_2 - d_2, \label{eq: bound1 on Interf. at R2 tighter CRC missing case with Z} \\
{\rm and} \quad f \big( ~ I(M_Y; ~ \mathbf{Z}_{[M_2+1:N_2]} ~  \big| M_Z, ~ \mathbf{H}) ~ \big) \leq N_2 - M_2. \label{eq: bound2 on Interf. at R2 tighter CRC missing case with Z}
\end{eqnarray}
\end{lemma}

Note that the bound in (\ref{eq: bound on Interf. at R2 preliminary CRC missing case}) can be recovered by simply adding inequalities (\ref{eq: bound1 on Interf. at R2 tighter CRC missing case with Z}) and (\ref{eq: bound2 on Interf. at R2 tighter CRC missing case with Z}), and therefore, these two bounds are tighter than the one in (\ref{eq: bound on Interf. at R2 preliminary CRC missing case}). Moreover, these bounds assert that if $d_2$ DoF are achievable for the second transmit-receive pair, then there exists an $M_2$-dimensional subspace (of the receive signal-space of R2), which carries interference with multiplexing gain at most $M_2 - d_2$. Thus, these bounds capture the notion of interference localization.

We now prove the above lemma.

\begin{IEEEproof}[Proof of Lemma \ref{lem: inter. localizn IC missing case}] Let $\tilde{Z}(t) \define \big( U^{22}(t) \big)^* Z(t)$. Since $\big( U^{22}(t)\big)^* H^{22}(t) = \Lambda^{22}(t) \big( V^{22}(t) \big)^*$, where $\Lambda^{22}(t)$ is diagonal with the bottom $N_2-M_2$ rows containing only zeros, we observe that the transmit signal $X^2(t)$ can not affect the last $N_2-M_2$ elements of $\tilde{Z}(t)$ $\forall$ $t$. In other words, $\mathbf{\tilde{Z}}_{[M_2+1:N_2]}$ is independent of $M_Z$, which yields
\begin{eqnarray}
\lefteqn{ \hspace{-0.3cm} ~ b R_2  \leq  I \big( M_Z ~ ; ~\mathbf{Z} ~ \big| ~\mathbf{H} \big) + b \epsilon_b = I \big( M_Z ~ ; ~ \mathbf{\tilde{Z}} ~ \big| ~ \mathbf{H} \big) + \epsilon_b } \nonumber  \\
&& {} \hspace{-0.8cm}  = I \big( M_Z ~; ~ \mathbf{\tilde{Z}}_{[1:M_2]} ~ \big| ~ \mathbf{\tilde{Z}}_{[M_2+1:N_2]}, ~ \mathbf{H} \big) + b \epsilon_b \nonumber  \\
&& {} \hspace{0.6cm} +  I \big( M_Z ~ ; ~ \mathbf{\tilde{Z}}_{[M_2+1:N_2]} ~ \big| ~ \mathbf{H} \big) \nonumber \\
&& {} \hspace{-0.8cm} = I \big( M_Z ~ ; ~ \mathbf{\tilde{Z}}_{[1:M_2]} ~ \big| ~ \mathbf{\tilde{Z}}_{[M_2+1:N_2]}, ~ \mathbf{H} \big) + b \epsilon_b \nonumber
\end{eqnarray}
Now, the techniques developed for deriving inequality (\ref{eq: bound on Interf. at R2 preliminary CRC missing case}) from (\ref{eq: bound on R2 missing case basic form}) can be used to obtain
\begin{eqnarray}
f \Big( I(M_Y; ~ \mathbf{\tilde{Z}}_{[1:M_2]} ~  \big| M_Z, ~ \mathbf{\tilde{Z}}_{[M_2+1:N_2]}, ~ \mathbf{H}) \Big) \leq M_2 - d_2. \label{eq: bound1 on Interf. at R2 tighter CRC missing case}
\end{eqnarray}
Moreover, by the single-user bound, we have
\begin{equation}
f \Big( I(M_Y; \mathbf{\tilde{Z}}_{[M_2+1:N_2]} ~  \big| M_Z, \mathbf{H}) \Big) \leq N_2 - M_2. \label{eq: bound2 on Interf. at R2 tighter CRC missing case}
\end{equation}

Note that conditioned on $M_Z$, $X^2(t)$ is deterministic. Since translation does not change differential entropy, it may be assumed that $X^2(t) = 0$ $\forall $ $t$ (see \cite[Proof of Lemma 2]{Vaze_Dof_final} for detailed proof). Thus, we may compute the mutual information terms in (\ref{eq: bound1 on Interf. at R2 tighter CRC missing case}) and (\ref{eq: bound2 on Interf. at R2 tighter CRC missing case}) by taking
\[
\tilde{Z}(t) = \big( U^{22}(t) \big)^* H^{21}(t) X^1(t) + \big( U^{22}(t)\big)^* W'(t).
\]
Further, note that $U^{22}(t)$, being a function of $H^{22}(t)$, is independent of $H^{21}(t)$. Since we have assumed here that $U^{21}(t) = I_{N_2}$ and $\Lambda^{21}(t) = I_{N_2}$, we have $ \big( U^{22}(t)\big)^* H^{21}(t) =  \big( U^{22}(t) \big)^* \big( V^{21}(t) \big)^*$. For any unitary matrix $U^{22}(t)$, it can be easily shown that $V^{21}(t) U^{22}(t)$ is still a semi-unitary matrix that is uniformly distributed over its domain. This implies that $V^{21}(t)$ is identically distributed as $V^{21}(t) U^{22}(t) $, which we denote symbolically as $V^{21}(t) \sim V^{21}(t) U^{22}(t)$. Moreover, $W'(t) \sim \big( U^{22}(t) \big)^* W'(t)$. Hence, conditioned on $M_Z$ and $\mathbf{H}$, $Z(t) \sim \tilde{Z}(t)$ or $\mathbf{Z} \sim \mathbf{\tilde{Z}}$. Hence, we get the lemma from bounds (\ref{eq: bound1 on Interf. at R2 tighter CRC missing case}) and (\ref{eq: bound2 on Interf. at R2 tighter CRC missing case}).
\end{IEEEproof}

\underline{Step III:} This is the final step of the analysis. Consider R1. Assuming that it knows message $M_Z$, we get via Fano's inequality that
\begin{eqnarray}
b R_1 & \leq & I(M_Y ; \mathbf{Y} | M_Z, \mathbf{H}) + b \epsilon_b \label{eq: bound on R1 missing case} \\
\Rightarrow d_1 & \leq &  f \big(  I(M_Y ; \mathbf{Y} | M_Z, \mathbf{H})  \big).  \label{eq: basic bound on d1 missing case}
\end{eqnarray}
Since conditioned on $M_Z$, $X_2(t)$ can be taken to be equal to $0$ $\forall$ $t$, $U^{i1}(t)$ has been assumed to be equal to $I_{N_1}$, and $\Lambda^{i1}(t)$ is diagonal, we observe that if $N_1 > M_1$, the last $N_1 - M_1$ antennas of R1 receive only noise at all times. Therefore, the random variables $\mathbf{Y}_{M_1+1:N_1}$ can be ignored in $f \big( I(M_Y ~ ; ~ \mathbf{Y} ~ \big| ~ M_Z, \mathbf{H}) \big)$, which upper-bounds $d_1$ in (\ref{eq: basic bound on d1 missing case}). Thus, henceforth in this section, it is assumed that $N_1 = N_1' = \min(N_1,M_1)$.

\begin{figure*}[!b]
\begin{picture}(10,1)
\put(10,10){\line(1,0){500}}
\end{picture}
\begin{eqnarray}
\bar{Y}(t) & \define & h^1_{\max}(t) \big( D^{11}(t) \big)^{-1} \big( U^{11}(t) \big)^* \begin{bmatrix} H^{11}(t) & H^{12}(t) \end{bmatrix} \begin{bmatrix} X^1(t) \\ X^2(t) \end{bmatrix} + W(t), \mbox{ where } D^{11}(t) = \begin{bmatrix} \tilde{\Lambda}^{11}(t) & 0_{p_1 \times q_1} \\ 0_{q_1 \times p_1} & I_{q_1 \times q_1} \end{bmatrix}. \! \! \label{eq: define Ybar for general isotropic case missing case}\\
\bar{Z}(t) & \define & h^1_{\max}(t) \big( \Lambda^{21}(t) \big)^{-1} \big( U^{21}(t) \big)^* \big\{ H^{21}(t) X^1(t) + H^{22}(t) X^2(t) \big\} + W'(t). \! \! \label{eq: define Zbar for general isotropic case missing case}
\end{eqnarray}
\begin{picture}(10,1)
\put(10,10){\line(1,0){500}}
\end{picture}
\begin{eqnarray}
\overline{\mathbf{D}}_{\rm{IC \backslash CRC}} \define \left\{ (d_1,d_2) \in \mathbb{R}^2_+ \Big| \forall ~ (w_1,w_2) \in \mathbb{R}^2_+, ~ w_1 d_1 + w_2 d_2 \leq \overline{\lim_{P \to \infty}} ~ \frac{1}{\log_2 P} \left[ \sup_{\big( R_1(P), R_2(P) \big) \in \mathcal{C}(P)} \Big\{ w_1 R_1 + w_2 R_2 \Big\} \right] \right\}. \label{eq: defn of the general DoF region missing case}
\end{eqnarray}
\end{figure*}

We now divide the antennas of R1 into two groups: the first group consists of the last $l \define N_2-M_2$ antennas of R1, while the second group contains the remaining  $N_1-l$ antennas. Then, using the chain rule for the mutual information \cite{CT}, we get
\begin{eqnarray}
\lefteqn{ d_1 \leq f \big(  I(M_Y ~ ; ~ \mathbf{Y} ~ \big| ~ M_Z, \mathbf{H})  \big) ~ \quad \mbox{ with } N_1 = N_1'} \nonumber \\
&& {} \hspace{-10pt}  =  f \big(  I(M_Y ~ ; ~ \mathbf{Y}_{[N_1-l+1:N_1]} ~ \big| ~ M_Z, ~ \mathbf{H})  \big) \label{eq: bound on d1 preliminary missing case}\\
&& {} + f \big( I(M_Y ~ ; ~ \mathbf{Y}_{[1: N_1-l]} ~ \big| ~ M_Z, ~ \mathbf{Y}_{[N_1-l+1:N_1]}, ~ \mathbf{H}) \big). \nonumber
\end{eqnarray}
We bound each of the two terms appearing in (\ref{eq: bound on d1 preliminary missing case}) starting with the first term. Toward this end, note that the isotropicity of the channel matrices and the assumptions made about $U^{i1}(t)$ and $\Lambda^{i1}(t)$ together imply that for any given $b$, the joint distribution of the random variables $\mathbf{Y}_{[N_1-l+1:N_1]}$, conditioned on $M_Z$ and $\mathbf{H}$, is identical to that of $\mathbf{Z}_{[M_2+1:N_2]}$, conditioned on $M_Z$ and $\mathbf{H}$. Hence,
\begin{eqnarray}
\lefteqn{ f \big( I(M_Y ~; ~ \mathbf{Y}_{[N_1-l+1:N_1]} ~ \big| ~ M_Z, ~ \mathbf{H})\big) } \nonumber \\
&& {} \hspace{-8pt} = f \big( I(M_Y; ~ \mathbf{Z}_{[M_2+1:N_2]} ~  \big| M_Z, ~ \mathbf{H}) \big) \leq N_2 - M_2. \label{eq: bound on first term of d1 missing case}
\end{eqnarray}
For the second term in (\ref{eq: bound on d1 preliminary missing case}), we have the following lemma.

\begin{lemma} \label{lem: main inequality missing case no CSIT DoF IC CRC}
If $l = N_2-M_2$, then
\begin{eqnarray}
\lefteqn{ f \big( I(M_Y ~ ; ~ \mathbf{Y}_{[1: N_1-l]} ~ \big| ~ M_Z, ~ \mathbf{Y}_{[N_1-l+1:N_1]}, ~ \mathbf{H}) \big) } \\
&& {} \hspace{-15pt} \leq \frac{N_1-l}{M_2} f \big( I(M_Y; ~ \mathbf{Z}_{[1:M_2]} ~  \big| M_Z, ~ \mathbf{Z}_{[M_2+1:N_2]}, ~ \mathbf{H}) \big) \nonumber \\
&& {} \hspace{-15pt} \leq \frac{N_1+M_2 - N_2}{M_2} (M_2 - d_2). \label{eq: ineq of lem: main inequality missing case no CSIT DoF IC CRC}
\end{eqnarray}
\end{lemma}
\begin{IEEEproof}
The last inequality follows from the definition of $l$ and inequality (\ref{eq: bound1 on Interf. at R2 tighter CRC missing case with Z}). Hence, it is sufficient to prove the first inequality, which is done in Appendix \ref{app: proof of lem: main inequality missing case no CSIT DoF IC CRC}.
\end{IEEEproof}

Substituting the inequalities (\ref{eq: bound on first term of d1 missing case}) and (\ref{eq: ineq of lem: main inequality missing case no CSIT DoF IC CRC}) into (\ref{eq: bound on d1 preliminary missing case}), we get
\begin{eqnarray*}
\lefteqn{ d_1 \leq N_2 - M_2 + \frac{N_1 + M_2 - N_2}{M_2} (M_2 - d_2) } \\
&& {} \Rightarrow d_1 + \frac{N_1 + M_2 - N_2}{M_2} d_2 \leq N_1,
\end{eqnarray*}
which is the desired inequality since $N_1 = N_1'$.

\emph{The general case without any assumptions about $U^{i1}(t)$ and $\Lambda^{i1}(t)$: } While this case follows from the techniques developed in \cite[Appendix D]{Vaze_Dof_final}, we include the details for the sake of completeness. We manipulate $Y(t)$ and $Z(t)$ to define $\bar{Y}(t)$ and $\bar{Z}(t)$ such that the mutual information terms in equations (\ref{eq: bound on R1 missing case}) and (\ref{eq: bound on R2 missing case}) are upper-bounded and the proof presented above holds if $\bar{Y}(t)$ and $\bar{Z}(t)$ are considered as the channel outputs. To this end, define $h(t)$ to be the maximum of all elements of matrices $\Lambda^{11}(t)$ and $\Lambda^{21}(t)$. Define $h^1_{\max}(t) = \max\{ 1, h(t)\}$. Further, $ p_1 \define \min(M_1,N_1)$, $q_1 \define N_1-p_1$. Recall that $\Lambda^{21}(t)$ is square to define $\bar{Y}(t)$ and $\bar{Z}(t)$ in equations (\ref{eq: define Ybar for general isotropic case missing case}) and (\ref{eq: define Zbar for general isotropic case missing case}) at the bottom the page. It can be proved that $I(M_Y;\mathbf{Y}|M_Z,\mathbf{H}) \leq I(M_Y;\mathbf{\bar{Y}}|M_Z,\mathbf{H}) $ and $I(M_Z;\mathbf{Z}|\mathbf{H}) \leq I(M_Z;\mathbf{\bar{Z}}|\mathbf{H}) $ (see proofs of Theorems 5 and 6 from \cite{Vaze_Dof_final}). Define $\bar{H}^{11}(t)$ and $\bar{H}^{12}(t)$ such that $\bar{Y}(t) = \bar{H}^{11}(t) X^1(t) + \bar{H}^{12}(t) X^2(t) + W(t)$ and analogously $\bar{H}^{21}(t)$ and $\bar{H}^{22}(t)$. Now the proof given above applies by making the following correspondence: $Y(t) \leftrightarrow \bar{Y}(t)$, $Z(t) \leftrightarrow \bar{Z}(t)$, and $H^{ij}(t) \leftrightarrow \bar{H}^{ij}(t)$ $\forall$ $i,j,t$.

\subsection{Comparison with the Proof of \cite{Zhu_Guo_noCSIT_DoF_2010}} \label{subsec: comparison with zhu guo}
Interference localization is central to the above proof as well as to the one in \cite{Zhu_Guo_noCSIT_DoF_2010}. However, the two works employ completely different techniques to prove this fact. In \cite{Zhu_Guo_noCSIT_DoF_2010}, the authors\footnote{In \cite{Zhu_Guo_noCSIT_DoF_2010}, the user ordering is exactly opposite of what is taken here.} first assume that R1 knows the message $M_Z$ (as we do here), and under this assumption, show that, as far as the DoF region is concerned, it is optimal for T2 to transmit $X^2(t)$ that is Gaussian with distribution $\mathcal{C}\mathcal{N}(0, \frac{P}{M_2}I_{M_2})$ (see Theorem 3 therein). Subsequently, for $X^2 \sim \mathcal{C}\mathcal{N}(0,\frac{P}{M_2}I_{M_2})$, it is proven using a lemma (namely, Lemma 4 therein) that the subspace spanned by $X^2(t)$ at R2 can not provide any information to it about the transmit signal $X^1(t)$ (cf. equation (47) therein), which in a way captures the interference localization phenomenon. In contrast, we prove here the same point in Lemma \ref{lem: inter. localizn IC missing case} using basic information-theoretic identities like the chain rule for mutual information \cite{CT}.

Another important step in our proof is Lemma \ref{lem: main inequality missing case no CSIT DoF IC CRC}, which again follows from simple identities such as conditioning reduces entropy, the chain rule for differential entropy, etc. On the other hand, the proof in \cite{Zhu_Guo_noCSIT_DoF_2010} needs a result (namely, Lemma 3 therein) that is a counterpart of Lemma \ref{lem: main inequality missing case no CSIT DoF IC CRC} we have here, its proof is given there using  more involved techniques which invoke the minimum mean squared error (MMSE). The proof here, in addition to be being simpler, is also more widely applicable, as we illustrate below.

We use the bounding techniques developed in this section to obtain the no-CSIT DoF region of the CRC with asymmetrically constrained transmitters (see Theorems \ref{thm: DoF region noCSIT CRC iid missing case} and \ref{thm: DoF region noCSIT CRC correlated missing case}) for which the inner and outer-bounds (to the no-CSIT DoF region) reported in \cite{Vaze_Dof_final} are not tight. However, the extension to this problem of the technique of \cite{Zhu_Guo_noCSIT_DoF_2010} is not known because their approach rests critically on being able to prove the optimality of choosing $X^2(t)$ to be Gaussian, which, in the context of the CRC, may not hold since T2 is now transmitting not just to R2 but also to R1.

Further, consider the problem of determining the generalized DoF (GDoF) region of the no-CSIT IC, where the GDoF region is defined to be equal to the DoF region when the gains (i.e., the Frobenius norm \cite{Horn-Johnson}) of the direct-link channel matrices ($H^{11}(t)$ and $H^{21}(t)$) and those of the cross-link channel matrices ($H^{12}(t)$ and $H^{22}(t)$) are unequal with the ratio of their values in dB equal to $\alpha \geq 0$ (the DoF region is the GDoF with $\alpha = 1$; see \cite{Etkin_Tse_Wang} for a formal definition). It turns out that for characterizing the no-CSIT GDoF region of the IC with asymmetrically constrained transmitters in the very weak interference regime of $\alpha \leq \frac{1}{2}$, it is necessary to prove that the interference is localized which, even in the more general setting of the GDoF analysis, can be done using the techniques developed above \cite{GDoF_region_noCSIT_IC_isit11}. In contrast, however, the applicability of the approach of \cite{Zhu_Guo_noCSIT_DoF_2010} is not clear (cf. \cite{GDoF_region_noCSIT_IC_isit11}). This is because for small values of $\alpha$, the bound obtained by assuming that R1 knows the message $M_Z$ is loose (since R1 at low $\alpha$ can not possibly decode $M_Z$, cf. \cite[Subsections III-C and III-D]{Etkin_Tse_Wang}). Hence, an outer-bound must be derived without assuming R1 to know the message $M_Z$, in which case the optimality of choosing $X^2(t)$ to be Gaussian (with a certain covariance matrix) can not be shown.

\subsection{The More General Definition}
\label{subsec: DoF region more general missing case}

As stated earlier, the definition of $\mathbf{D}_{\rm{IC \backslash CRC}}$ in \eqref{eq:defn-dof} is restrictive. Here we define the DoF region (cf. \cite{JaferShamai}) more generally (by relaxing the requirement that the limits $\lim_{P \to \infty} \frac{R_i(P)}{\log_2 P}$, $i =1,2$ exist) to be the region $\overline{\mathbf{D}}_{\rm{IC \backslash CRC}}$ in equation \eqref{eq: defn of the general DoF region missing case} at the bottom of the previous page, where $\mathbb{R}_+$ denotes the set of non-negative real numbers. Comparing the two definitions, we have
\[
\mathbf{D}_{\rm{IC \backslash CRC}} \subseteq \overline{\mathbf{D}}_{\rm{IC \backslash CRC}}
\]

The techniques developed in the earlier part of this section allow us to characterize $\overline{\mathbf{D}}_{\rm{IC}}$ as per the following theorem.
\begin{theorem} \label{cor: more general DoF region IC}
With no CSIT, we have for the MIMO IC with isotropic fading and for the MIMO CRC with i.i.d. (or correlated) Rayleigh fading, we have
\[
\overline{\mathbf{D}}_{\rm{IC \backslash CRC}} = \mathbf{D}_{\rm{IC \backslash CRC}}
\]
\end{theorem}
\begin{IEEEproof}
For the MIMO IC, it is sufficient to prove that if $(d_1,d_2) \in \overline{\mathbf{D}}_{\rm{IC}}$, then the bound $L$
\[
d_1 + \frac{N_1' + M_2 - N_2}{M_2} d_2 \leq N_1'
\]
holds with $N_1' = \min(M_1,N_1)$. See Appendix \ref{app: proof of cor: more general DoF region IC}. The DoF region $\overline{\mathbf{D}}_{\rm{CRC}}$ can be shown to coincide with $\mathbf{D}_{\rm{CRC}}$ determined in Theorems \ref{thm: DoF region noCSIT CRC iid missing case} and \ref{thm: DoF region noCSIT CRC correlated missing case} for i.i.d. and correlated Rayleigh fading cases, respectively, in an analogous manner.
\end{IEEEproof}

\section{Proof of Theorem \ref{thm: DoF region noCSIT CRC iid missing case}: $L$ is an Outer-Bound} \label{sec: proof of thm: DoF region noCSIT CRC iid missing case}

The goal of this section is to show that for the MIMO CRC with asymmetrically constrained transmitters and i.i.d. Rayleigh fading, bound $L$ is an outer-bound with $N_1' = \min(M_1 + M_2,N_1)$. In the following sub-section, we first deal with the case of $N_1 \geq M_1 + M_2$; later, in Section \ref{subsec: proof of thm: DoF region noCSIT CRC iid missing case N1 < M1 + M2}, we address the remaining case of $N_1 < M_1 + M_2$.

\subsection{Case of $N_1 \geq M_1 + M_2$}

The proof again consists of three steps, as was the case in the last section. At Step I, we bound $d_1$. At Step II, derive the interference localization property; and at Step III, bound $d_1$.

\underline{Step I:} Fano's inequality yields us
\begin{equation}
bR_2 \leq I \big( M_Z;\mathbf{Z} \big| \mathbf{H} \big) + b \epsilon_b.
\end{equation}

\underline{Step II:} To prove that the interference is localized at R2, we make use of the following lemma, which gives us the QR-decomposition \cite{Horn-Johnson} of $H^{22}(t)$.
\begin{lemma} \label{lem: QR  decomposn missing case}
An i.i.d. Rayleigh-faded $N_2 \times M_2$ channel matrix $H^{22}(t)$ can be written as
\[
H^{22}(t) = Q^{22}(t) R^{22}(t),
\]
where $Q^{22}(t)$ and $R^{22}(t)$ are deterministic functions of $H^{22}(t)$ such that
\begin{enumerate}[(i)]
\item $Q^{22}(t)$ is an $N_2 \times N_2$ isotropically-distributed unitary matrix;
\item $R^{22}(t)$ is an $N_2 \times M_2$ upper-triangular matrix, i.e., the $M_2 \times M_2$ square matrix formed by retaining just the first $M_2$ rows of it is upper-triangular, while the bottom $N_2 - M_2$ rows of it consist only of zeros;
\item entries of $R^{22}(t)$, which are not surely zero, follow a continuous distribution (i.e., their cumulative distribution function is continuous and differentiable); and
\item all entries of $R^{22}(t)$ are independent of each other and also of the unitary matrix $Q^{22}(t)$. \end{enumerate}
\end{lemma}
\begin{IEEEproof}
Follows from the definition of QR-decomposition \cite{Horn-Johnson} and \cite[Lemma 2.1]{TulinoVerdu}.
\end{IEEEproof}

With $Q^{22}(t)$ and $R^{22}(t)$ obtained as per the above lemma, define
\begin{equation}
\tilde{Z}(t) \define \big( Q^{22}(t) \big)^* Z(t). \label{eq: defn of Ztilde CRC missing case}
\end{equation}
This construction allows us to obtain the following lemma.
\begin{lemma}[Interference Localization]
The following bounds hold:
\begin{eqnarray}
\hspace{-0.5cm} f \big( I(M_Y; \mathbf{\tilde{Z}}_{[1:M_2]} ~  \big| M_Z, \mathbf{\tilde{Z}}_{[M_2+1:N_2]}, \mathbf{H}) \big) \hspace{-7pt} & \leq & \hspace{-7pt} M_2 - d_2, \label{eq: bound1 on Interf. at R2 tighter CRC(actually) missing case} \\
\mbox{and} ~ f \big( I(M_Y; \mathbf{\tilde{Z}}_{[M_2+1:N_2]} ~  \big| M_Z, \mathbf{H}) \big) \hspace{-7pt} & \leq & \hspace{-7pt} N_2 - M_2. \label{eq: bound2 on Interf. at R2 tighter CRC(actually) missing case}
\end{eqnarray}
\end{lemma}
\begin{IEEEproof}
Since the bottom $N_2 - M_2$ entries of $R^{22}(t)$ consist only of zeros, the transmit signal $X^2(t)$ can not affect the last $N_2 - M_2$ entries of $\tilde{Z}(t)$. Moreover, the signal $X^1(t)$ is independent of $M_Z$. This observation and the analysis in the proof of Lemma \ref{lem: inter. localizn IC missing case} up to equations (\ref{eq: bound1 on Interf. at R2 tighter CRC missing case}) and (\ref{eq: bound2 on Interf. at R2 tighter CRC missing case}) allow us to derive the inequalities of the lemma.
\end{IEEEproof}

Recall, in the previous section, we were able to claim that the above bounds hold even with $\mathbf{\tilde{Z}}$ replaced by $\mathbf{Z}$ because in the case of the IC, conditioned on $M_Z$, $X^2(t)$ can be taken to be deterministic. However, this need not be the case with the CRC where T2 knows both the messages. As a result, in the present case, the above bounds do not hold with $\mathbf{\tilde{Z}}$ replaced by $\mathbf{Z}$. This necessitates a more sophisticated analysis at Step III for the CRC.

Before proceeding further, we state a corollary which simplifies the computation of the mutual information terms appearing in the above two equations.
\begin{corollary} \label{cor: structure of Ztilde CRC missing case}
In the mutual information terms appearing in inequalities (\ref{eq: bound1 on Interf. at R2 tighter CRC(actually) missing case}) and (\ref{eq: bound2 on Interf. at R2 tighter CRC(actually) missing case}), it may be assumed that $\mathbf{\tilde{Z}} = \big\{ \tilde{Z}(t) \big\}_{t=1}^b$, where
\begin{equation}
\tilde{Z}(t) = H^{21}(t) X^1(t) + R^{22}(t) X^2(t) + W'(t). \label{eq: cor: structure of Ztilde CRC missing case}
\end{equation}
\end{corollary}
\begin{IEEEproof}
See Appendix \ref{sub-app: proof of cor: structure of Ztilde CRC missing case}.
\end{IEEEproof}
Thus, henceforth, we assume that equation (\ref{eq: cor: structure of Ztilde CRC missing case}) holds and $\tilde{Z}(t)$ is treated as the signal received by R2.

\underline{Step III:} Consider now R1. Assuming that it knows $M_Z$, we get via Fano's inequality that
\begin{equation}
d_1 \leq f \big( I ( M_Y; \mathbf{Y} \big| M_Z, \mathbf{H}) \big).
\end{equation}
Suppose $N_1 > M_1 + M_2$. Then, at any given $t$, R1 can construct a noisy version of the channel inputs $X^1(t)$ and $X^2(t)$ using just $M_1+M_2$ channel outputs $Y_{[1:M_1+M_2]}(t)$ by inverting a matrix
\[
\begin{bmatrix} H^{11}_{[1:M_1+M_2]}(t) & H^{12}_{[1:M_1+M_2]}(t) \end{bmatrix},
\]
(which can be done with probability $1$). Hence, the last $N_1 - (M_1 + M_2)$ channel outputs at R1 can not contribute to the DoF of the CRC, and therefore, they can be ignored in the present analysis (see \cite[Section II-C]{Vaze_Dof_final} for detailed proof of this claim). It is thus assumed in this sub-section that $N_1 \leq M_1 + M_2$ and $N_1' = N_1$.

We would like to use the inequalities (\ref{eq: bound1 on Interf. at R2 tighter CRC(actually) missing case}) and (\ref{eq: bound2 on Interf. at R2 tighter CRC(actually) missing case}) to bound the term $f \big( I( M_Y;\mathbf{Y} \big| M_Z, \mathbf{H}) \big)$. However, the channel matrices $H^{11}(t)$ and $H^{12}(t)$ corresponding to R1 are i.i.d. Rayleigh faded, while those corresponding to R2 (which observes $\tilde{Z}(t)$) are not (see equation (\ref{eq: cor: structure of Ztilde CRC missing case})). As a result, the inequalities (\ref{eq: bound1 on Interf. at R2 tighter CRC(actually) missing case}) and (\ref{eq: bound2 on Interf. at R2 tighter CRC(actually) missing case}) can not directly be used to bound $f \big( I( M_Y;\mathbf{Y} \big| M_Z, \mathbf{H}) \big)$. Instead, we first need to manipulate this term to bring it to a form that is suitable for the application of bounds in (\ref{eq: bound1 on Interf. at R2 tighter CRC(actually) missing case}) and (\ref{eq: bound2 on Interf. at R2 tighter CRC(actually) missing case}).

The analysis henceforth is divided into four steps, namely, Steps III.a - III.d. Before getting into the details of these steps, we explain below the outline since the analysis is complicated. This outline has also been depicted in Table \ref{table: outline of step III missing case CRC(actually)} in the context of the CRC with $(M_1,M_2,N_1,N_2) = (5,2,7,3)$.

$\bullet$ {\bf Step III.a: } At this step, we upper-bound the mutual information term $I( M_Y;\mathbf{Y} \big| M_Z, \mathbf{H})$ by assuming that R1, at time $t$, observes not just the actual channel outputs $Y_i(t)$, $i \in [1:N_1]$, but also some extra fictitious channel outputs which are defined shortly. See Step III.a in Table \ref{table: outline of step III missing case CRC(actually)}). The fictitious outputs are added such that we have $N_2$ outputs $Y_{S_i}$ corresponding to each set $S_i$ and $Y_{S_i}(t) \sim Z(t)$.

$\bullet$ {\bf Step III.b: } Here, we use the QR-decomposition of Lemma \ref{lem: QR  decomposn missing case} to transform the outputs $Y_{S_i}(t)$ into $\tilde{Y}_{S_i}(t)$ such that $\tilde{Y}_{S_i}(t) \sim \tilde{Z}(t)$.

$\bullet$ {\bf Step III.c: } It is shown that the upper-bound on $d_1$ obtained at Step III.b can be tightened by suitably removing some of the entries of $\tilde{Y}_{S_i}$. See Step III.c in Table \ref{table: outline of step III missing case CRC(actually)}.

$\bullet$ {\bf Step III.d: } This is the final step at which the bounds (\ref{eq: bound1 on Interf. at R2 tighter CRC(actually) missing case}), (\ref{eq: bound2 on Interf. at R2 tighter CRC(actually) missing case}), and the one obtained at Step III.c are used to derive the desired bound $L$.

We now proceed to the proof. \vspace{1mm}

\begin{table*}[H]
To begin, we bound $d_1$ via Fano's inequality as
\[
d_1 \leq f \Big( ~ I \big( M_Y; \mathbf{Y} \Big| M_Z, \mathbf{H} \big) ~ \Big).
\]
The proof now proceeds through the following four steps. \vspace{2mm}

{\bf Step III.a} : Group the actual channel outputs into $4$ sets and add  fictitious channel outputs so that each set contain $N_2 =3$ outputs.

\begin{centering}
\subfloat{ {\renewcommand{\arraystretch}{1.5} \begin{tabular}{|c|c|c|}
\hline
\multicolumn{1}{|c|}{Set} & indices & \multicolumn{1}{c|}{Actual Outputs} \tabularnewline
\hline
\hline
$S_{1}^{a}$ & $\{1\}$ & $Y_{1}$\tabularnewline
\hline
$S_{2}^{a}$ & $\{2\}$ & $Y_{2}$\tabularnewline
\hline
$S_{3}^{a}$ & $\{3,4\}$ & $Y_{3}, \: Y_{4}$\tabularnewline
\hline
$S_{0}^{a}$ & $\{5,6,7\}$ & $Y_{5},\: Y_{6},\: Y_{7}$\tabularnewline
\hline
\end{tabular} } }\hspace{1cm}\subfloat{ { \renewcommand{\arraystretch}{1.5} \begin{tabular}{|c|c|c|}
\hline
Set & indices & Fictitious outputs \tabularnewline
\hline \hline
$S_{1}^{f}$ & $\{f11,\: f12\}$ & $Y_{f11},\: Y_{f12}$\tabularnewline
\hline
$S_{2}^{f}$ & $\{f21,\: f22\}$ & $Y_{f21},\: Y_{f22}$\tabularnewline
\hline
$S_{3}^{f}$ & $\{f31\}$ & $Y_{f31}$\tabularnewline
\hline
$S_{0}^{f}$ & $\{\phi\}$ & $\phi$\tabularnewline
\hline
\end{tabular} } }
\par\end{centering}

\begin{centering}
\vspace{1mm}\subfloat{{ \renewcommand{\arraystretch}{1.8} \centering{}\begin{tabular}{|c|c|c|}
\hline
Set & indices & All Outputs \tabularnewline
\hline \hline
$S_{1}=S_{1}^{a}\cup S_{1}^{f}$ & $\{1,\: f11,\: f12\}$ & $Y_{S_{1}}=\{Y_{1},\: Y_{f11},\: Y_{f12}\}$\tabularnewline
\hline
$S_{2}=S_{2}^{a}\cup S_{2}^{f}$ & $\{2,\: f21, \: f22\}$ & $Y_{S_{2}}=\{Y_{2},\: Y_{f21},\: Y_{f22}\}$\tabularnewline
\hline
$S_{3}=S_{3}^{a}\cup S_{3}^{f}$ & $\{3,\: 4, \: f31\}$ & $Y_{S_{3}}=\{Y_{3},\: Y_{4},\: Y_{f31}\}$\tabularnewline
\hline
$S_{0}=S_{0}^{a}\cup S_{0}^{f}$ & $\{5,\: 6, \: 7\}$ & $Y_{S_{0}}=\{Y_{5},\: Y_{6},\: Y_{7}\}$\tabularnewline
\hline
\end{tabular}}}
\par\end{centering}
\begin{centering}
\vspace{1.5mm}
\par\end{centering}

Note that $Y_{S_i} \sim Z$ $\forall$ $i$. After adding these fictitious outputs, we get
\[
d_1 \leq f \Big( I \big( M_Y; \mathbf{Y}_{S_1}, \, \mathbf{Y}_{S_2}, \, \mathbf{Y}_{S_3}, \, \mathbf{Y}_{S_0} \Big| M_Z, \mathbf{H} \big) \Big). 
\]

{\bf Step III.b} : Use the procedure that allows us to define $\tilde{Z}$ from $Z$ to define $\tilde{Y}_{S_i}$ from $Y_{S_i}$.

\begin{centering}
\subfloat{{ \renewcommand{\arraystretch}{1.8} \centering{}\begin{tabular}{|c|c|}
\hline
Set & Outputs after transformation \tabularnewline
\hline \hline
$S_{1}$ & $\tilde{Y}_{S_{1}}=\left\{ \tilde{Y}_{S_{1}1},\: \tilde{Y}_{S_{1}2},\: \tilde{Y}_{S_{1}3}\right\} $ \tabularnewline
\hline
$S_{2}$ & $\tilde{Y}_{S_{2}}=\left\{ \tilde{Y}_{S_{2}1},\: \tilde{Y}_{S_{2}2},\: \tilde{Y}_{S_{2}3}\right\} $ \tabularnewline
\hline
$S_{3}$ & $\tilde{Y}_{S_{3}}=\left\{ \tilde{Y}_{S_{3}1},\: \tilde{Y}_{S_{3}2},\: \tilde{Y}_{S_{3}3}\right\} $ \tabularnewline
\hline
$S_{0}$ & $\tilde{Y}_{S_{0}}=\left\{ \tilde{Y}_{S_{0}1},\: \tilde{Y}_{S_{0}2},\: \tilde{Y}_{S_{0}3}\right\} $ \tabularnewline
\hline
\end{tabular}} }
\par\end{centering}
\vspace{1.5mm}
Note that $\tilde{Y}_{S_i} \sim \tilde{Z}$ $\forall$ $i$. After this transformation, we obtain
\[
d_1 \leq f \Big( ~ I \big( M_Y; \mathbf{\tilde{Y}}_{S_1}, \, \mathbf{\tilde{Y}}_{S_2}, \, \mathbf{\tilde{Y}}_{S_3}, \, \mathbf{\tilde{Y}}_{S_0} \Big| M_Z, \mathbf{H}, \mathbf{H}_f \big) ~ \Big).
\]

{\bf Step III.c} : Retain few entries of $\tilde{Y}_{S_{i}},$ $i=1,\,2,\,3,\,4.$

$\mathcal{P}(1) \define \left\{ (1,1), \:(2,1), \:(3,1), \:(3,2), \:(0,1), \:(0,2), \:(0,3)\right\} .$ $\mathcal{P}(2) \define \left\{ (1,2), \:(2,2), \:(3,1), \:(3,2), \:(0,1), \:(0,2),\:(0,3)\right\} .$

\begin{centering} \subfloat{ { \renewcommand{\arraystretch}{2} \begin{tabular}{|c|c|}
\hline
Set & Outputs\tabularnewline
\hline
\hline
$\mathcal{P}(1)$ & $\tilde{Y}_{\mathcal{P}(1)}=\left\{ \tilde{Y}_{S_{1}1},\ \tilde{Y}_{S_{2}1},\ \tilde{Y}_{S_{3}1},\ \tilde{Y}_{S_{3}2},\ \tilde{Y}_{S_{0}1},\ \tilde{Y}_{S_{0}2},\ \tilde{Y}_{S_{0}3}\right\} $\tabularnewline
\hline
$\mathcal{P}(2)$ & $\tilde{Y}_{\mathcal{P}(2)}=\left\{ \tilde{Y}_{S_{1}2},\ \tilde{Y}_{S_{2}2},\ \tilde{Y}_{S_{3}1},\ \tilde{Y}_{S_{3}2},\ \tilde{Y}_{S_{0}1},\ \tilde{Y}_{S_{0}2},\ \tilde{Y}_{S_{0}3}\right\} $\tabularnewline
\hline
\end{tabular} } }
\par \end{centering} \vspace{1.5mm}

It is proved that
\[
d_1 \leq \frac{1}{2} f \Big( I \big( M_Y; \mathbf{\tilde{Y}}_{\mathcal{P}(1)} \Big| M_Z, \mathbf{H}, \mathbf{H}_f \big) \Big) +  \frac{1}{2} f \Big( I \big( M_Y; \mathbf{\tilde{Y}}_{\mathcal{P}(2)} \Big| M_Z, \mathbf{H}, \mathbf{H}_f \big) \Big).
\]

{\bf Step III.d } : The right hand side of the above equation is bounded through inequalities in (\ref{eq: bound1 on Interf. at R2 tighter CRC(actually) missing case}) and (\ref{eq: bound2 on Interf. at R2 tighter CRC(actually) missing case}). \vspace{4mm}

\caption{Outline of Step III for the i.i.d. Rayleigh-faded CRC. $(M_1,M_2,N_1,N_2) = (5,2,7,3)$} \label{table: outline of step III missing case CRC(actually)}
\end{table*}

$\bullet$ \underline{Step III.a:}

Before adding the fictitious channel outputs, we group the actual channel outputs at R1 into a certain number of sets. Then corresponding to each set, we add some fictitious outputs so that we have in total $N_2$ outputs corresponding to each set (see Table \ref{table: outline of step III missing case CRC(actually)}).

Toward this end, we first introduce some terminology. Let
\[
m \define \left\lfloor ~ \frac{N_1 - N_2}{M_2} ~ \right\rfloor \mbox{ and } n \define (N_1 - N_2) - m M_2 \geq 0,
\]
where $\lfloor x \rfloor$ denotes the largest integer that is less than or equal to $x$. We now partition the set $[1:N_1]$ into $m+n+1$ disjoint subsets as follows:
\begin{eqnarray*}
S^a_i & \define & \{i\} ~ ~ \forall ~ i \in [1:n], \\
S^a_{n+j} & \define & [n + (j-1)M_2 + 1 : n + j M_2] ~ ~ \forall ~ j \in [1:m],\\
S^a_0 & \define & [N_1-N_2+1:N_1].
\end{eqnarray*}
We now define the {\em fictitious} channel outputs as follows: For an $l \in [0:n+m]$ and a $k \in [1:N_2 -1]$, define
\[
Y_{flk}(t) = H^{11}_{flk}(t) X^1(t) + H^{12}_{flk}(t) X^2(t) + w_{flk}(t),
\]
where $H^{11}_{flk}(t) \in \mathbb{C}^{1 \times M_1}$ and $H^{12}_{flk}(t) \in \mathbb{C}^{1 \times M_2}$ are fictitious channel vectors; $w_{flk}(t) \in \mathbb{C}$ is a fictitious noise variable; and the entries of $H^{11}_{flk}(t)$, $H^{12}_{flk}(t)$, and $w_{flk}(t)$ are i.i.d. $\mathcal{C}\mathcal{N}(0,1)$ random variables, which are also i.i.d. across $l$, $k$, and $t$, and are also independent of actual channel matrices $\{H^{ij}(t)\}_{i,j,t}$ and the actual noises $\{W(t),W'(t)\}_t$. Moreover, the transmitters are unaware of the realizations of the fictitious channel vectors and the fictitious noises, while the receivers know the realizations of the fictitious channel vectors.

Now, define
\begin{eqnarray*}
S^f_i & \hspace{-5pt} \define \hspace{-5pt} & \big\{fi1,fi2,\cdots,fi(N_2-1) \big\}, ~ \forall i \in [1:n] \\
S^f_j & \hspace{-5pt} \define \hspace{-5pt} & \big\{ fj1, fj2, \cdots, fj(N_2-M_2) \big\}, ~ \forall j \in [n+1:n+m], \\
S^f_0 & \hspace{-5pt} \define \hspace{-5pt} & \phi \mbox{ (the empty set)} \\
S_i & \hspace{-5pt} \define \hspace{-5pt} &  S_i^a \cup S_i^f ~ \forall ~ i \in [0:n+m].
\end{eqnarray*}
Thus, the cardinality of $S_i$ is $N_2$ (denoted symbolically as $\big| S_i \big| = N_2$) $\forall$ $i \in [0:n+m]$. For each set $S_i$, define
\[
Y_{S_i}(t) = \Big\{Y_k(t) \Big\}_{k \in S_i^a} \cup \Big\{ Y_{fij}(t) \Big\}_{fij \in S_i^f}
\]
and $ \mathbf{Y}_{S_i} \define \left\{ Y_{S_i}(t) \right\}_{t=1}^b$. Let $H_f(t)$ be the collection of the realizations of the fictitious channel matrices at time $t$ and let $\mathbf{H}_f = \big\{ H_f(t) \big\}_{t=1}^b$, then we have
\begin{equation}
d_1 \leq f \Big( ~ I \big(M_Y; ~ \big\{ \mathbf{Y}_{S_i} \big\}_{i=0}^{n+m} ~ \big| ~ M_Z, \mathbf{H}, \mathbf{H}_f \big) ~ \Big).
\end{equation}

The following corollary allows us to determine the distribution of $Y_{S_i}(t)$.
\begin{corollary} \label{cor: structure of Y_Si CRC missing case}
Given an $i \in [0:n+m]$, we may write
\begin{eqnarray}
Y_{S_i}(t) = H^{11}_{S_i}(t) \cdot X^1(t) + H^{12}_{S_i}(t) \cdot X^2(t) + W_{S_i}(t), \label{eq: cor: structure of Y_Si CRC missing case}
\end{eqnarray}
for some $H^{11}_{S_i}(t) \in \mathbb{C}^{N_1 \times M_1}$, $H^{12}_{S_i}(t) \in \mathbb{C}^{N_1 \times M_2}$, and $W_{S_i}(t) \in \mathbb{C}^{N_1 \times 1}$ such that their entries follow the $\mathcal{C}\mathcal{N}(0,1)$ distribution and are i.i.d. among themselves and also across $i$ and $t$. Hence, for any $i, j \in [0:n+m]$ with $i \not= j$, we have
\[
Y_{S_i}(t) \sim Y_{S_j}(t) \sim Z(t)
\]
with $Y_{S_i}(t)$ and $Y_{S_j}(t)$ being independent.
\end{corollary}
\begin{IEEEproof}
See Appendix \ref{sub-app: proof of cor: structure of Y_Si CRC missing case}.
\end{IEEEproof}
Thus, the outputs corresponding to each set are identically distributed as $Z(t)$. This serves as the basis for the further manipulations. \vspace{1mm}

$\bullet$ \underline{Step III.b:}

In this step, we use the QR-decomposition introduced in Lemma \ref{lem: QR  decomposn missing case} to transform the channel outputs $Y_{S_i}(t)$ to $\tilde{Y}_{S_i}(t)$ such that $\tilde{Y}_{S_i}(t) \sim \tilde{Z}(t)$ $\forall$ $t,i$.

Using the QR-decomposition of Lemma \ref{lem: QR  decomposn missing case}, we write
\[
H^{12}_{S_i}(t) = Q^{12}_{S_i}(t) \cdot R^{12}_{S_i}(t),
\]
where $Q^{12}_{S_i}(t)$ is an $N_2 \times N_2$ unitary matrix, $R^{12}_{S_i}(t)$ is an $N_2 \times M_2$ upper-triangular matrix, and they satisfy properties (i)-(iv) of Lemma \ref{lem: QR  decomposn missing case}. Define
\begin{eqnarray}
\tilde{Y}_{S_i}(t) & = & \begin{bmatrix} \tilde{Y}_{S_i 1}(t) & \tilde{Y}_{S_i 2}(t) & \cdots & \tilde{Y}_{S_i N_2}(t) \end{bmatrix}^T \\
& \define & \big( Q^{12}_{S_i}(t) \big)^* \cdot Y_{S_i}(t) \label{eq: def Ytilde_Si CRC missing case} \\
\mbox{and } \mathbf{\tilde{Y}}_{S_i} & \define & \Big\{ \tilde{Y}_{S_i} (t) \Big\}_{t=1}^{b}.
\end{eqnarray}
Note that $\tilde{Y}_{S_i}(t)$ consists of $N_2$ entries. Since a unitary operation can not alter mutual information, we get
\begin{equation}
d_1 \leq f \Big( I \big(M_Y; ~ \big\{ \mathbf{\tilde{Y}}_{S_i} \big\}_{i=0}^{n+m} ~ \big| ~ M_Z, \mathbf{H}, \mathbf{H}_f \big) \Big). \label{eq: conclusion of Step II CRC missing case thm2}
\end{equation}

We have the following corollary which shows that $\tilde{Y}_{S_i}(t) \sim \tilde{Z}(t)$.
\begin{corollary} \label{cor: structure of Ytilde_Si CRC missing case}
In the mutual information term in (\ref{eq: conclusion of Step II CRC missing case thm2}), we may write
\begin{equation}
\tilde{Y}_{S_i}(t) = H^{11}_{S_i}(t) \cdot X^1(t) + R^{12}_{S_i}(t) \cdot X^2(t) + W_{S_i}(t).
\end{equation}
Moreover, for any $i,j \in [0:n+m]$ with $i \not= j$, we have
\[
\tilde{Y}_{S_i}(t) \sim \tilde{Y}_{S_j}(t) \sim \tilde{Z}(t)
\]
with $\tilde{Y}_{S_i}(t)$ being independent of $\tilde{Y}_{S_j}(t)$.
\end{corollary}
\begin{IEEEproof}
Follows from Corollaries \ref{cor: structure of Ztilde CRC missing case} and \ref{cor: structure of Y_Si CRC missing case}.
\end{IEEEproof} \vspace{1mm}
Thus, the outputs $\tilde{Y}_{S_i}(t)$ have the same distribution as that of $\tilde{Z}(t)$, which, as we will see shortly, is important for being able to use the bounds in (\ref{eq: bound1 on Interf. at R2 tighter CRC(actually) missing case}) and (\ref{eq: bound2 on Interf. at R2 tighter CRC(actually) missing case}). \vspace{1mm}

$\bullet$ \underline{Step III.c:}

At this step, we alter the mutual information term appearing in (\ref{eq: conclusion of Step II CRC missing case thm2}) by retaining just $\big| S_i^a \big|$ entries of $\tilde{Y}_{S_i}(t)$ for each $t$. To this end, consider the following. Let
\begin{eqnarray}
\mathcal{P} & \define &  \Big\{ (i,j) \Big| ~ i \in [0:n+m], ~ j \in [1:N_2] \Big\} \mbox{ and} \nonumber \\
\tilde{Y}_{\mathcal{P}}(t) & \define & \Big\{ \tilde{Y}_{S_i j}(t) \Big\}_{(i,j) \in \mathcal{P}} \mbox{ so that} \nonumber \\
\mathbf{\tilde{Y}}_{\mathcal{P}} & \define & \Big\{ \tilde{Y}_{\mathcal{P}}(t) \Big\}_{t=1}^b = \Big\{ \mathbf{\tilde{Y}}_{S_i} \Big\}_{i=0}^{n+m} \mbox{ and} \nonumber \\
d_1 & \leq & f \Big( I \big(M_Y; ~ \mathbf{\tilde{Y}}_{\mathcal{P}} ~ \big| ~ M_Z, \mathbf{H}, \mathbf{H}_f \big) \Big). \label{eq: conclusion of Step II diff form CRC missing case thm2}
\end{eqnarray}

Now, consider $M_2$ subsets, $\mathcal{P}(k)$, $k \in [1:M_2]$, of $\mathcal{P}$, which are defined as follows. For a given $k \in [1:M_2]$, $\mathcal{P}(k)$ is defined as the set of all ordered pairs $(i,j)$ for which
\begin{eqnarray*}
j \in  \begin{cases} \{k\}, & \mbox{if } i \in [1:n], \\
[1:M_2], & \mbox{if } i \in [n+1:n+m], \\
[1:N_2], & \mbox{if } i =0. \end{cases}
\end{eqnarray*}
Note that the cardinality of each of the above sets is $N_1$. Let $\mathcal{P}^c(k) = \mathcal{P} \backslash \mathcal{P}(k)$. Define
\begin{equation}
\tilde{Y}_{\mathcal{P}(k)}(t) = \Big\{ \tilde{Y}_{S_i j}(t) \Big\}_{(i,j) \in \mathcal{P}(k)} \mbox{ and } \mathbf{\tilde{Y}}_{\mathcal{P}(k)} = \Big\{ \tilde{Y}_{\mathcal{P}(k)}(t) \Big\}_{t=1}^b \label{eq: defn of Ytilde_Pk missing case CRC}
\end{equation}
and analogously $\tilde{Y}_{\mathcal{P}^c(k)}(t)$ and $\mathbf{\tilde{Y}}_{\mathcal{P}^c(k)}$.

Then, using the inequality (\ref{eq: conclusion of Step II diff form CRC missing case thm2}), we get the following:
\begin{eqnarray}
\lefteqn{ \hspace{-10pt} d_1 \leq  \frac{1}{M_2} \sum_{k=1}^{M_2} \biggl\{ f \Big( I \big(M_Y; ~ \mathbf{\tilde{Y}}_{\mathcal{P}(k)} ~ \big| M_Z, \mathbf{H}, \mathbf{H}_f \big) \Big) \biggr. } \nonumber \\
&& {} \hspace{13pt} \biggl. + f \Big(  I \big(M_Y; ~ \mathbf{\tilde{Y}}_{ \mathcal{P}^c(k) } ~  \big| M_Z, \mathbf{H}, \mathbf{H}_f, ~ \mathbf{\tilde{Y}}_{\mathcal{P}(k)} \big) ~ \Big) \biggr\} \nonumber \\
&& {} \hspace{-23pt} = \frac{1}{M_2} \sum_{k=1}^{M_2} f \Big(  I \big(M_Y; \mathbf{\tilde{Y}}_{\mathcal{P}(k)} \big| M_Z, \mathbf{H}, \mathbf{H}_f \big) \Big), \label{eq: conclusion of Step III CRC thm2 missing case} \\
&& {} \hspace{-23pt} = \frac{1}{M_2} \sum_{k=1}^{M_2} f \Big(  h \big( \mathbf{\tilde{Y}}_{\mathcal{P}(k)} \big| M_Z, \mathbf{H}, \mathbf{H}_f \big) \Big) , \label{eq: conclusion-new of Step III CRC thm2 missing case}
\end{eqnarray}
where the equality (\ref{eq: conclusion of Step III CRC thm2 missing case}) follows due to the Lemma \ref{lem: full-rankness CRC thm2 missing case} below; and the last equality is a simple application of Lemma \ref{lem: in app mutual_info=entropy in f} proved in Appendix \ref{app: proof of lem: main inequality missing case no CSIT DoF IC CRC}. We now state and prove Lemma \ref{lem: full-rankness CRC thm2 missing case}.
\begin{lemma} \label{lem: full-rankness CRC thm2 missing case}
For any given $k \in [1:M_2]$, we have
\[
f \Big(  I \big(M_Y; ~ \mathbf{\tilde{Y}}_{ \mathcal{P}^c(k) } ~  \big| M_Z, \mathbf{H}, \mathbf{H}_f, ~ \mathbf{\tilde{Y}}_{\mathcal{P}(k)} ~ \big) \Big) = 0.
\]
\end{lemma}
\begin{IEEEproof}
See Appendix \ref{app: proof of lem: full-rankness CRC thm2 missing case}.
\end{IEEEproof}
This step thus allows us to tighten the bound derived at Step III.b. \vspace{1mm}

$\bullet$ \underline{Step III.d:}

This is the last step. Here, we bound the differential entropy term appearing in equation (\ref{eq: conclusion-new of Step III CRC thm2 missing case}) via bounds in (\ref{eq: bound1 on Interf. at R2 tighter CRC(actually) missing case}) and (\ref{eq: bound2 on Interf. at R2 tighter CRC(actually) missing case}) to derive finally the desired bound $L$. In the following, we denote by $\mathcal{V}$ the collection $\{M_Z, \mathbf{H}, \mathbf{H}_f\}$ and we also need the following lemma.
\begin{lemma}  \label{lem: eq: Step IV.4 CRC thm2 missing case}
For a given $i \in [0:n+m]$ and a $k \in [1:M_2]$, the joint distribution of random variables
\[
\Big\{ M_Y, M_Z, \mathbf{H}, \mathbf{H}_f, \mathbf{\tilde{Y}}_{S_i [1:k]}, \mathbf{\tilde{Y}}_{S_0 [k+1:N_2]} \Big\}
\]
is identical to that of the random variables
\[
\Big\{ M_Y, M_Z, \mathbf{H}, \mathbf{H}_f, \mathbf{\tilde{Z}}_{[1:k]}, \mathbf{\tilde{Z}}_{[k+1:N_2]}  \Big\}.
\]
\end{lemma}
\begin{IEEEproof}
See Appendix \ref{app: proof of lem: eq: Step IV.4 CRC thm2 missing case}.
\end{IEEEproof}

From inequality (\ref{eq: conclusion-new of Step III CRC thm2 missing case}), we get the following:
\begin{eqnarray}
d_1 & \hspace{-6pt}  \leq  \hspace{-5pt} & \frac{1}{M_2} \sum_{k=1}^{M_2} \biggl\{ f \Big( h \big( \mathbf{\tilde{Y}}_{S_0 [1:N_2]} \big| \mathcal{V} \big) \Big) \biggr. \nonumber \\
& & ~ ~ ~  + \sum_{i = 1}^{n}  f \Big( h \big( \mathbf{\tilde{Y}}_{S_i k} \big| \mathcal{V}, \mathbf{\tilde{Y}}_{S_0 [1:N_2]} \big) \Big) \nonumber  \\
& & ~ ~ ~  + \biggl. \sum_{j=n+1}^{n+m}  f \Big( h \big( \mathbf{\tilde{Y}}_{S_j [1:M_2]} \big| \mathcal{V}, \mathbf{\tilde{Y}}_{S_0 [1:N_2]} \big) \Big) \biggr\} \label{eq: Step IV.1 CRC thm2 missing case}\\
& \hspace{-6pt}  = \hspace{-5pt} & f \Big( h \big( \mathbf{\tilde{Y}}_{S_0 [1:N_2]} \big| \mathcal{V} \big) \Big) \nonumber  \\
& & ~ ~ ~  + \sum_{i=1}^{n} \frac{1}{M_2} \sum_{k=1}^{M_2} f \Big( h \big( \mathbf{\tilde{Y}}_{S_i k} \big| \mathcal{V}, \mathbf{\tilde{Y}}_{S_0 [1:N_2]} \big) \Big) \nonumber  \\
& & ~ ~ ~  + \sum_{j=n+1}^{n+m} f \Big( h \big( \mathbf{\tilde{Y}}_{S_j [1:M_2]} \big| \mathcal{V}, \mathbf{\tilde{Y}}_{S_0 [1:N_2]} \big) \Big) \label{eq: Step IV.2 CRC thm2 missing case} \\
& \hspace{-6pt}  \leq  \hspace{-5pt} & f \Big( h \big( \mathbf{\tilde{Y}}_{S_0 [1:N_2]} \big| \mathcal{V} \big) \Big) \nonumber  \\
& & + \sum_{i=1}^{n} \frac{1}{M_2} \sum_{k=1}^{M_2} f \Big( h \big( \mathbf{\tilde{Y}}_{S_i k} \big| \mathcal{V}, \mathbf{\tilde{Y}}_{S_0 [k+1:N_2]} \big) \Big) \nonumber  \\
& & + \sum_{j=n+1}^{n+m} f \Big( h \big( \mathbf{\tilde{Y}}_{S_j [1:M_2]} \big| \mathcal{V}, \mathbf{\tilde{Y}}_{S_0 [M_2+1:N_2]} \big) \Big) \label{eq: Step IV.3 CRC thm2 missing case} \\
& \hspace{-6pt}  =  \hspace{-5pt} & f \Big( h \big( \mathbf{\tilde{Z}} \big| \mathcal{V} \big) \Big) \nonumber  \\
& & ~ ~ ~ + \sum_{i=1}^{n} \frac{1}{M_2} \sum_{k=1}^{M_2} f \Big( h \big( \mathbf{\tilde{Z}}_k \big| \mathcal{V}, \mathbf{\tilde{Z}}_{[k+1:N_2]} \big) \Big) \nonumber  \\
& & ~ ~ ~ + \sum_{j=n+1}^{n+m} f \Big( h \big( \mathbf{\tilde{Z}}_{[1:M_2]} \big| \mathcal{V}, \mathbf{\tilde{Z}}_{[M_2+1:N_2]} \big) \Big) \label{eq: Step IV.4 CRC thm2 missing case} \\
& \hspace{-6pt}  =  \hspace{-5pt} & f \Big( h \big( \mathbf{\tilde{Z}} \big| \mathcal{V} \big) \Big) \nonumber  \\
& &  ~ ~ ~ + \sum_{i=1}^{n} \frac{1}{M_2} f \Big( h \big( \mathbf{\tilde{Z}}_{[1:M_2]} \big| \mathcal{V}, \mathbf{\tilde{Z}}_{[M_2+1:N_2]} \big) \Big) \nonumber  \\
& & ~ ~ ~  + \sum_{j=n+1}^{n+m} f \Big( h \big( \mathbf{\tilde{Z}}_{[1:M_2]} \big| \mathcal{V}, \mathbf{\tilde{Z}}_{[M_2+1:N_2]} \big) \Big). \label{eq: Step IV.5 CRC thm2 missing case} \\
& \hspace{-6pt}  =  \hspace{-5pt} & f \Big( I \big( M_Y; \mathbf{\tilde{Z}} \big| M_Z, \mathbf{H} \big) \Big) \nonumber  \\
& & + \sum_{i=1}^{n} \frac{1}{M_2} f \Big( I \big( M_Y; \mathbf{\tilde{Z}}_{[1:M_2]} \big| M_Z, \mathbf{H}, \mathbf{\tilde{Z}}_{[M_2+1:N_2]} \big) \Big) \nonumber  \\
& & + \sum_{j=n+1}^{n+m} f \Big( I \big( M_Y; \mathbf{\tilde{Z}}_{[1:M_2]} \big| M_Z, \mathbf{H}, \mathbf{\tilde{Z}}_{[M_2+1:N_2]} \big) \Big) \label{eq: Step IV.6 CRC thm2 missing case}
\end{eqnarray}
where (\ref{eq: Step IV.1 CRC thm2 missing case}) follows due to the chain rule for the differential entropy and due to the fact that conditioning reduces entropy \cite{CT}, the inequality (\ref{eq: Step IV.3 CRC thm2 missing case}) holds since conditioning reduces differential entropy, the next equality (\ref{eq: Step IV.4 CRC thm2 missing case}) is true because of Lemma \ref{lem: eq: Step IV.4 CRC thm2 missing case} stated earlier, equality in (\ref{eq: Step IV.5 CRC thm2 missing case}) follows by the chain rule for the differential entropy and the last equality holds since $\mathbf{H}_f$ is independent of all other random variables.

Now, bounds in (\ref{eq: bound1 on Interf. at R2 tighter CRC(actually) missing case}), (\ref{eq: bound2 on Interf. at R2 tighter CRC(actually) missing case}), and (\ref{eq: Step IV.6 CRC thm2 missing case}), and the fact that $n + m M_2 = N_1 - N_2$ together yield
\begin{eqnarray*}
\lefteqn{ d_1 \leq  (N_2 - M_2) + (M_2 - d_2) \left\{ \frac{n}{M_2} + m  \right\} }\\
&& {} \hspace{-10pt} \Rightarrow d_1 \leq  N_2 - d_2 + (M_2 - d_2) \frac{N_1-N_2}{M_2} \\
&& {} \hspace{-10pt} \Rightarrow d_1 + d_2 \frac{N_1 + M_2 - N_2}{M_2} \leq N_1,
\end{eqnarray*}
which is the desired bound $L$.

\subsection{Case of $N_1 < M_1 + M_2$} \label{subsec: proof of thm: DoF region noCSIT CRC iid missing case N1 < M1 + M2}

To derive the DoF region of the no-CSIT CRC with $M_1 + M_2 > N_1 > N_2 > M_2$, we prove that the DoF region of the given $(M_1,M_2,N_1,N_2)$ CRC is equal to that of the $(N_1-M_2,M_2,N_1,N_2)$ CRC, which has been derived in the earlier subsection. Hence, the result follows. Toward this end, we will manipulate the input-output relationship of the given $(M_1,M_2,N_1,N_2)$ CRC in such a manner that it resembles that of the $(N_1-M_2,M_2,N_1,N_2)$ CRC with i.i.d. Rayleigh fading, whose DoF region is known from the previous analysis of the previous subsection.

With this motivation, let
\[
H^i(t) \define \begin{bmatrix} H^{i1}(t) & H^{i2}(t) \end{bmatrix}, ~ i \in \{1, 2\} \mbox{ and } X(t) = \begin{bmatrix} X^1(t) \\ X^2(t) \end{bmatrix}
\]
so that
\[
Y(t) = H^1(t) X(t) + W(t) \mbox{ and } Z(t) = H^2(t) X(t) + W'(t).
\]
Moreover, let $\mathbf{H^i} = \big\{ H^i(t) \big\}_{t=1}^b$.

Fano's inequality yields
\begin{eqnarray*}
b R_2 & \leq &  I (M_Z; \mathbf{Z} \big| \mathbf{H}) + b \epsilon_b, \mbox{ and} \\
b R_1 & \leq & I (M_Y; \mathbf{Y} \big| M_Z, \mathbf{H}) + b \epsilon_b. \\
\end{eqnarray*}
Now, note that conditioned on $\mathbf{H^2}$ ($\mathbf{H^1}$), random variables $M_Z$, $M_Y$, and $\mathbf{Z}$ ($\mathbf{Y}$) are independent of $\mathbf{H^1}$ ($\mathbf{H^2}$). Hence, we get
\begin{eqnarray*}
b R_2 & \leq & I (M_Z; \mathbf{Z} \big| \mathbf{H^2}) + b \epsilon_b, \mbox{ and} \label{eq: step0 bound on d2 CRC-B missing case}\\
b R_1 & \leq & I (M_Y; \mathbf{Y} \big| M_Z, \mathbf{H^1}) + b \epsilon_b. \label{eq: step0 bound on d1 CRC-B missing case} \\
\end{eqnarray*}

Consider now the following lemma which yields the singular-value decomposition of $H^i(t)$.
\begin{lemma} \label{lem: singular-value decomposn CRC-B missing case}
For a given $i \in \{1,2\}$, an $N_i \times (M_1 + M_2)$ i.i.d. Rayleigh-faded channel matrix $H^i(t)$ can be written as
\[
H^i(t) = U^i(t) \begin{bmatrix} \Lambda^i(t) & 0_{N_i \times n_i} \end{bmatrix} \big( U^{i1}(t) \big)^*,
\]
where matrices $U^i(t)$, $\Lambda^i(t)$, and $U^{i1}(t)$ are deterministic functions of $H^i(t)$ such that
\begin{enumerate}[(i)]
\item $U^i(t)$ is an $N_i \times N_i$ isotropically-distributed unitary matrix,
\item $\Lambda^i(t)$ is an $N_i \times N_i$ diagonal matrix with non-negative diagonal entries,
\item $0_{N_i \times n_i}$ is an $N_i \times n_i$ all-zero matrix with $n_i = M_1 + M_2 - N_i$,
\item $U^{i1}(t)$ is an $(M_1 + M_2) \times (M_1 + M_2)$ isotropically-distributed matrix and
\item $U^i(t)$, $\Lambda^i(t)$, and $U^{i1}(t)$ are independent of each other.
\end{enumerate}
\end{lemma}
\begin{IEEEproof}
Follows from the definition of the singular-value decomposition \cite{Horn-Johnson} and \cite[Lemma 2.6, Example 2.6]{TulinoVerdu}.
\end{IEEEproof}

Thus, if $V^i(t)$ denotes the semi-unitary matrix obtained by retaining just the first $N_1$ columns of $U^{i1}(t)$, then we may write
\begin{eqnarray}
H^1(t) & = & U^1(t) \Lambda^1(t) \big( V^1(t) \big)^* \mbox{ and} \\
H^2(t) & = & U^2(t) \begin{bmatrix} \Lambda^2(t) & 0_{N_1 \times (n_2-n_1)} \end{bmatrix} \big( V^2(t) \big)^*
\end{eqnarray}
with $V^1(t) \sim V^2(t)$. This implies that the mutual information terms in (\ref{eq: step0 bound on d2 CRC-B missing case}) and (\ref{eq: step0 bound on d1 CRC-B missing case}) remain unaffected, even if we assume that $V^1(t) = V^2(t) \define V(t)$.

Since $V(t)$ is uniformly distributed over the set of semi-unitary matrices, we have
\[
V(t) \sim V(t) \big( Q^i(t) \big)^*,
\]
where $Q^1(t)$ and $Q^2(t)$ are $N_1 \times N_1$ isotropically-distributed unitary matrices that are independent of each other and all other random variables, and also independent across $t$. Hence, we get
\begin{eqnarray*}
H^1(t) & \sim & U^1(t) \Lambda^1(t) \big( Q^1(t) \big)^* \big( V(t) \big)^* \mbox{ and} \label{eq: introducing Q H1 CRC-B missing case} \\
H^2(t) & \sim & U^2(t) \begin{bmatrix} \Lambda^2(t) & 0_{N_1 \times (n_2-n_1)} \end{bmatrix} \big( Q^2(t) \big)^* \big( V(t) \big)^*. \label{eq: introducing Q H2 CRC-B missing case}
\end{eqnarray*}
Hence, it may be assumed that the above two equations hold even with `$\sim$' replaced by equality `$=$'. Now, we introduce some terminology:
\begin{eqnarray*}
\overline{H}^1(t) & \define & U^1(t) \Lambda^1(t) \big( Q^1(t) \big)^*, \\
\overline{H}^2(t) & \define & U^2(t) \begin{bmatrix} \Lambda^2(t) & 0_{N_1 \times (n_2-n_1)} \end{bmatrix} \big( Q^2(t) \big)^*, \\
\overline{X}(t) & \define & \big( V(t) \big)^* X(t), \\
\overline{Y}(t) & \define & \overline{H}^1(t) \overline{X}(t) + W(t), \mbox{ and}\\
\overline{Z}(t) & \define & \overline{H}^2(t) \overline{X}(t) + W'(t). \label{eq: overlines introduced CRC-B missing case}
\end{eqnarray*}
Then it is not difficult to see that
\begin{eqnarray}
b R_2 & \leq & I (M_Z; \mathbf{\overline{Z}} \big| \mathbf{H^2}, \mathbf{Q^2}) + b \epsilon_b , \mbox{ and} \label{eq: step-overline bound on d2 CRC-B missing case}\\
b R_1 & \leq & I (M_Y; \mathbf{\overline{Y}} \big| M_Z, \mathbf{H^1}, \mathbf{Q^1}) + b \epsilon_b . \label{eq: step-overline bound on d1 CRC-B missing case}
\end{eqnarray}

Note here that $||\overline{X}(t)||^2 \leq ||X^1(t)||^2 + ||X^2(t)||^2$, which implies that
\begin{equation}
\lim_{b \to \infty} \frac{1}{b} \sum_{t=1}^b \mathbb{E} ||\overline{X}(t)||^2 \leq 2P . \label{eq: power constraint with overline X CRC-B missing case}
\end{equation}
This fact will be used later.

Note that the signal $X^1(t)$ is independent of $M_Z$, while $X^2(t)$, which is $M_2$-dimensional, is dependent on it.

Consider an $N_1 \times N_1$ unitary matrix $E(t)$ such that the span of the last $M_2$ columns of it is equal to the span of last $M_2$ columns of $\big( V(t) \big)^*$. Then define
\begin{equation}
\overline{\overline{X}}(t) = \big( E(t) \big)^* \overline{X}(t). \label{eq: defn overline overline X CRC-B missing case}
\end{equation}
The following corollary helps in determining its distribution.
\begin{corollary} \label{cor: structure of overline overline X CRC-B missing case}
The first $N_1 - M_2$ entries of $\overline{\overline{X}}(t)$ are independent of $M_Z$ $\forall$ $t$.
\end{corollary}
\begin{IEEEproof}
See Appendix \ref{sub-app: proof of cor: structure of overline overline X CRC-B missing case}.
\end{IEEEproof}
Note that $E(t)$ is a function of $V(t)$, and thus, is independent of $Q^1(t)$ and $Q^2(t)$, which yields $\big(Q^i(t)\big)^* E(t) \sim \big( Q^i(t) \big)^*$. This implies that
\[
H^i(t) E(t) \sim H^i(t), ~ i=1,2
\]
and moreover $H^i(t) E(t)$ is independent of $E(t)$ and hence of $V(t)$. Therefore, from (\ref{eq: overlines introduced CRC-B missing case}) and (\ref{eq: defn overline overline X CRC-B missing case}), we have
\begin{eqnarray*}
\overline{Y}(t) & \sim & \overline{\overline{Y}}(t) \define \overline{H}^1(t) \overline{\overline{X}}(t) + W(t) \mbox{ and} \\
\overline{Z}(t) & \sim & \overline{\overline{Z}}(t) \define \overline{H}^2(t) \overline{\overline{X}}(t) + W'(t). \\
\end{eqnarray*}
Hence, we have
\begin{eqnarray}
b R_2 & \leq & I (M_Z; \mathbf{\overline{\overline{Z}}} \big| \mathbf{H^2}, \mathbf{Q^2}) + b \epsilon_b , \mbox{ and} \label{eq: step-overline bound on d2 CRC-B missing case}\\
b R_1 & \leq & I (M_Y; \mathbf{\overline{\overline{Y}}} \big| M_Z, \mathbf{H^1}, \mathbf{Q^1}) + b \epsilon_b. \label{eq: step-overline bound on d1 CRC-B missing case}
\end{eqnarray}

Note that by Lemma \ref{lem: singular-value decomposn CRC-B missing case}, channel matrices $\overline{H}^1(t)$ and $\overline{H}^2(t)$ are i.i.d. Rayleigh faded (see their definitions). Moreover, they are also independent across $t$, and independent of $\overline{\overline{X}}(t)$ and of additive noises. Further, $\overline{\overline{X}}(t)$ satisfies the power constraint of $2P$ and the first $N_1-M_2$ entries of it are independent of $M_Z$. Therefore, if the tuple $(R_1, R_2)$ is such that $(bR_1,bR_2)$ satisfies bounds (\ref{eq: step-overline bound on d2 CRC-B missing case}) and (\ref{eq: step-overline bound on d1 CRC-B missing case}), then the analysis of the previous sub-section (which is general enough to address the case of power constraint being $2P$) performed by making a correspondence that $M_1 \leftrightarrow N_1-M_2$, $M_2 \leftrightarrow M_2$, $N_1 \leftrightarrow N_1$, and $N_2 \leftrightarrow N_2$ implies that $(d_1, d_2)$ must satisfy the inequality
\[
d_1 + d_2 \frac{N_1 + M_2-N_2}{M_2} \leq N_1,
\]
which coincides with the desired bound $L$ since $N_1 = N_1' = \min\{M_1+M_2,N_1\}$.

\section{Proof of Theorem \ref{thm: DoF region noCSIT CRC correlated missing case}: $L$ is an Outer-Bound} \label{sec: proof of thm: DoF region noCSIT CRC correlated missing case}

The proof of $L$ being an outer-bound follows exactly along the lines of that presented in Section \ref{sec: proof of thm: DoF region noCSIT IC missing case} with some appropriate modifications. We present here an outline.

We have
\begin{eqnarray*}
d_2 & \leq & f \big( I (M_Z;\mathbf{Z} \big| \mathbf{H}) \big) \mbox{ and} \\
d_1 & \leq & f \big( I (M_Y;\mathbf{Y} \big| M_Z, \mathbf{H}) \big).
\end{eqnarray*}
As argued in Section \ref{sec: proof of thm: DoF region noCSIT CRC iid missing case}, we may assume without loss of generality that $N_1 = N_1' \leq M_1 + M_2$.

Define
\[
\tilde{Y}(t) = \big( U^{12} \big)^*Y(t) \mbox{ and } \tilde{Z}(t) = \big( U^{22} \big)^* Z(t).
\]
Then it can be easily shown that
\begin{eqnarray*}
f \big( I(M_Y; ~ \mathbf{\tilde{Z}}_{[1:M_2]} ~  \big| M_Z, ~ \mathbf{\tilde{Z}}_{[M_2+1:N_2]}, ~ \mathbf{H}) \big) & \leq & M_2 - d_2. \\
f \big( I(M_Y; \mathbf{\tilde{Z}}_{[M_2+1:N_2]} ~  \big| M_Z, \mathbf{H}) \big) & \leq & N_2 - M_2.
\end{eqnarray*}
Now the analysis in Section \ref{sec: proof of thm: DoF region noCSIT IC missing case} from equation (\ref{eq: basic bound on d1 missing case}) onwards holds with $\mathbf{Y}$ replaced by $\mathbf{\tilde{Y}}$ (with $N_1 = N_1' = \min(N_1,M_1+M_2)$) and $\mathbf{Z}$ replaced by $\mathbf{\tilde{Z}}$. The desired bound $L$ can thus be derived.

\section{Conclusion} \label{sec: conclusion missing case}
A simpler and more generic (and hence more widely applicable) proof is given than the one found recently in \cite{Zhu_Guo_noCSIT_DoF_2010} of the DoF region of the MIMO IC with $\min(M_1,N_1) > M_2 > N_2$. This proof is based on the idea of interference localization. Using this idea, the exact DoF region of the MIMO CRC with $\min(M_1+M_2,N_1) > N_2 > M_2$ is also characterized for which the bounds proposed earlier in \cite{Vaze_Dof_final} were not tight.

\appendices

\section{Proof of Lemma \ref{lem: main inequality missing case no CSIT DoF IC CRC}}
\label{app: proof of lem: main inequality missing case no CSIT DoF IC CRC}

This proof is identical in principle to the proof of \cite[Lemma 1]{Vaze_Dof_final}.
Let $\mathcal{A} = \{M_Z, \mathbf{Y}_{[N_1-l+1:N_1]}, \mathbf{H}\}$ and $\mathcal{B} = \{M_Z, \mathbf{Z}_{[M_2+1:N_2]}, \mathbf{H} \}$. First, consider the following lemma.
\begin{lemma} \label{lem: in app mutual_info=entropy in f}
We have
\[
f \Big( I \big( M_Y; \mathbf{Z}_{[1:M_2]}  \big| \mathcal{B} \big) \Big) =  f \Big( h (\mathbf{Z}_{[1:M_2]}  \big| \mathcal{B} \big) \Big).
\]
\end{lemma}
\begin{IEEEproof}
Using the definition of mutual information, we obtain
\begin{eqnarray}
\lefteqn{ f \Big( I \big( M_Y; \mathbf{Z}_{[1:M_2]}  \big| \mathcal{B} \big) \Big) } \nonumber \\
&& {} = f \Big( h \big( \mathbf{Z}_{[1:M_2]}  \big| \mathcal{B} \big) -  h \big( \mathbf{Z}_{[1:M_2]}  \big| M_Y, \mathcal{B} \big) \Big) \nonumber \\
&& {} = f \Big( h \big( \mathbf{Z}_{[1:M_2]}  \big| \mathcal{B} \big) - h \big( \mathbf{W}'_{[1:M_2]} \big) \Big) \label{eq: step2 lem: in app mutual_info=entropy in f}\\
&& {} = f \big( h (\mathbf{Z}_{[1:M_2]}  \big| \mathcal{B}) \big), \label{eq: step3 lem: in app mutual_info=entropy in f}
\end{eqnarray}
where equality in (\ref{eq: step2 lem: in app mutual_info=entropy in f}) holds since \begin{inparaenum}[(a)] \item conditioned on $M_Y$ and $M_Z$, transmit signals are deterministic, \item translation does not change differential entropy, and \item noise is independent of channel matrices and messages; \end{inparaenum} while the last equality (\ref{eq: step3 lem: in app mutual_info=entropy in f}) is true because $\lim_{P \to \infty} \frac{1}{\log_2 P} \Big\{ \lim_{b \to \infty} \frac{1}{b} h (\mathbf{W}'_{[1:M_2]}) \Big\} = 0$, which follows from the following facts: \begin{inparaenum} \item noise random variables are i.i.d. across time and receive antennas according $\mathcal{C}\mathcal{N}(0,1)$ distribution, and therefore, \item $h (\mathbf{W}'_{[1:M_2]}) = b \cdot o(\log_2 P)$, where $o(\log_2 P)$ represents a term that is constant with $b$ such that $\lim_{P \to \infty} \frac{o(\log_2 P)}{\log_2 P} = 0$. \end{inparaenum}
\end{IEEEproof}

Applying the above lemma, we observe that the desired inequality holds provided the inequality
\begin{eqnarray}
(N_1 - l) f \big( h(\mathbf{Z}_{[1:M_2]}  \big| \mathcal{B}) \big) \geq M_2 f \big( h(\mathbf{Y}_{[1: N_1-l]} \big| \mathcal{A}) \big) \label{eq: proof of main lemma missing case sufficient condition}
\end{eqnarray}
is true. The goal of the remainder of this appendix is to prove the above inequality. To this end, consider two sets of random variables $\mathcal{Z} = \{\mathbf{Z}_1, \mathbf{Z}_2, \cdots, \mathbf{Z}_{M_2} \}$ and $\mathcal{Y} = \{ \mathbf{Y}_1, \mathbf{Y}_2, \cdots, \mathbf{Y}_{N_1-l}\}$. In the following discussion, we treat $\mathbf{Z}_1$ as one random variable (although it is a random vector) and similarly the others. Then by symmetry of the distribution of the fading channel matrices, we get the following. For an integer $m$ such that $0 < m \leq \min(M_2,N_1-l)$, the joint distribution of any $m$ (distinct) random variables chosen from set $\mathcal{Z}$, when conditioned on $\mathcal{B}$, is identical to that of any $m$ (distinct) random variables chosen from set $\mathcal{Y}$, when conditioned on $\mathcal{A}$. Moreover, due to the same reason, for integer $m$ such that $0 < m \leq N_1-l$, the joint distribution of any $m$ (distinct) random variables chosen from the set $\mathcal{Y}$, when conditioned on $\mathcal{A}$, would be the same, regardless of which $m$ random variables are chosen. These facts yield
\begin{eqnarray}
f \big( ~ h(\mathbf{Z}_{[1:M_2]} ~  \big| \mathcal{B}) ~ \big) = f \big( ~ h( \mathbf{Y}_{[N_1-N_2+1: N_1-l]} ~ \big| \mathcal{A}) ~ \big) \label{eq: proof of main lemma missing case step 1}
\end{eqnarray}
Suppose the following is true: $(N_1-N_2) f \big( h(\mathbf{Z}_{[1:M_2]}  \big| \mathcal{B}) ~ \big)$
\begin{eqnarray}
\geq M_2 f \big( h( \mathbf{Y}_{[1: N_1-N_2]} ~ \big| \mathcal{A}, ~ \mathbf{Y}_{[N_1-N_2+1: N_1-l]}) \big). \label{eq: proof of main lemma missing case step 2}
\end{eqnarray}
Then we can add $M_2$ times equation (\ref{eq: proof of main lemma missing case step 1}) to the above inequality to obtain the required inequality (\ref{eq: proof of main lemma missing case sufficient condition}) (recall, $l = N_2 - M_2$), which shows the sufficiency of  proving the inequality (\ref{eq: proof of main lemma missing case step 2}). Now, the equality of conditional joint distributions discussed above, the chain rule for differential entropy, and the fact that conditioning reduces entropy together imply that
\begin{eqnarray*}
\lefteqn{ (N_1-N_2) \cdot f \big( ~ h(\mathbf{Z}_{[1:M_2]} ~  \big| \mathcal{B}) ~ \big)} \\
&& {} \hspace{-15pt} \geq M_2 (N_1-N_2) \cdot f \big( ~ h(\mathbf{Z}_1 ~  \big| \mathbf{Z}_{[2:M_2]}, ~ \mathcal{B}) ~ \big)  \\
&& {} \hspace{-15pt} \geq M_2 (N_1-N_2) f \big( ~ h( \mathbf{Y}_{(N_1-N_2)} ~ \big| \mathcal{A}, ~ \mathbf{Y}_{[N_1-N_2+2: N_1-l]}) ~ \big) \\
&& {} \hspace{-15pt} \geq M_2 (N_1-N_2) f \big( ~ h( \mathbf{Y}_{(N_1-N_2)} ~ \big| \mathcal{A}, ~ \mathbf{Y}_{[N_1-N_2+1: N_1-l]}) ~ \big) \\
&& {} \hspace{-15pt} \geq M_2  f \big( ~ h( \mathbf{Y}_{[1: N_1-N_2]} ~ \big| \mathcal{A}, ~ \mathbf{Y}_{[N_1-N_2+1: N_1-l]}) ~ \big),
\end{eqnarray*}
which yields the sought inequality and hence the lemma (cf. \cite[equations (14) and (15)]{Vaze_Dof_final}).

\section{Proof of Theorem \ref{cor: more general DoF region IC}} \label{app: proof of cor: more general DoF region IC}

Recall that the analysis of Section \ref{sec: proof of thm: DoF region noCSIT IC missing case} consists of two parts. In the first part, certain assumptions regarding the distribution of $U^{i1}(t)$ and $\Lambda^{i1}(t)$ are made, which are relaxed in the second part (towards the end) of the analysis. From the discussion therein, it is clear that, here, without loss of generality, we may restrict ourselves to the special case considered in part one. Accordingly, in the following, we let $U^{11}(t) = I_{N_1}$, $U^{21}(t) = I_{N_2}$, and all singular values of $H^{11}(t)$ and $H^{21}(t)$ are equal to unity with probability $1$.

As before, the proof consists of three steps. The main idea of the proof is identical to the one present in Section \ref{sec: proof of thm: DoF region noCSIT IC missing case}. We point out just the differences. \newline $\bullet$ \underline{Step I:} Applying Fano's inequality, we obtain
\begin{equation}
bR_2 \leq I \big( M_Z; \mathbf{Z} \big| \mathbf{H} \big) + b \epsilon_b.
\end{equation}
$\bullet$ \underline{Step II:} As done in the proof of Lemma \ref{lem: inter. localizn IC missing case}, we obtain
\begin{eqnarray*}
b R_2 & \leq &  I \big( M_Z ~ ; ~ \mathbf{\tilde{Z}}_{[1:M_2]} ~ \big| ~ \mathbf{\tilde{Z}}_{[M_2+1:N_2]}, ~ \mathbf{H} \big) + b \epsilon_b \\
& = & I \big( M_Y, M_Z ~ ; ~ \mathbf{\tilde{Z}}_{[1:M_2]} ~ \big| ~ \mathbf{\tilde{Z}}_{[M_2+1:N_2]}, ~ \mathbf{H} \big) + b \epsilon_b \\
& & \hspace{10pt} - I \big( M_Y ~ ; ~ \mathbf{\tilde{Z}}_{[1:M_2]} ~ \big| ~ M_Z, \mathbf{\tilde{Z}}_{[M_2+1:N_2]}, ~ \mathbf{H} \big).
\end{eqnarray*}
After equation (\ref{eq: bound2 on Interf. at R2 tighter CRC missing case}), it is shown in the proof of Lemma \ref{lem: inter. localizn IC missing case} that conditioned on $M_Z$ and $\mathbf{H}$, $\mathbf{Z} \sim \mathbf{\tilde{Z}}$. The same set of arguments and the last equation together yield
\begin{eqnarray}
\lefteqn{ I \big( M_Y ~ ; ~ \mathbf{Z}_{[1:M_2]} ~ \big| ~ M_Z, \mathbf{Z}_{[M_2+1:N_2]}, ~ \mathbf{H} \big) } \\
&& {} \hspace{-2mm} \leq I \big( M_Y, M_Z ~ ; ~ \mathbf{\tilde{Z}}_{[1:M_2]} ~ \big| ~ \mathbf{\tilde{Z}}_{[M_2+1:N_2]}, ~ \mathbf{H} \big) - b R_2 + b \epsilon_b. \label{eq: tight bound on inter. at R2 IC missing case general DoF}
\end{eqnarray}
This inequality serves as a counterpart of (\ref{eq: bound1 on Interf. at R2 tighter CRC missing case with Z}). Moreover, nothing that can be viewed as a counterpart of (\ref{eq: bound2 on Interf. at R2 tighter CRC missing case with Z}) is needed in the present case. \newline $\bullet$ \underline{Step III:} By denoting $N_1' = \min(M_1,N_1)$ and by applying Fano's inequality, we have
\begin{eqnarray*}
b R_1 & \leq & I(M_Y ; \mathbf{Y} | M_Z, \mathbf{H}) + b \epsilon_b \\
& = & I(M_Y; \mathbf{Y}_{[1:N_1']} | M_Z, \mathbf{H}) + b \epsilon_b \\
& & \hspace{15pt} + I (M_Y;\mathbf{Y}_{[N_1'+1:N_1]} | M_Z, \mathbf{H}, \mathbf{Y}_{[1:N_1']}).
\end{eqnarray*}
It is argued in Section \ref{sec: proof of thm: DoF region noCSIT IC missing case} after equation (\ref{eq: basic bound on d1 missing case}) that the last $N_1-N_1'$ antennas at R1 do not contribute to the DoF when it knows the message $M_Z$. Similar, arguments allow us to show that
\[
I (M_Y;\mathbf{Y}_{[N_1'+1:N_1]} | M_Z, \mathbf{H}, \mathbf{Y}_{[1:N_1']}) = b \cdot o(\log_2 P) ,
\]
where $o(\log_2 P)$ is constant with $b$ (cf. \cite[Lemma 3]{Vaze_Varanasi_delay_MIMO_IC}). Hence, for $l = N_2- M_2$, we have
\begin{eqnarray}
b R_1 & \leq & I(M_Y; \mathbf{Y}_{[1:N_1']} | M_Z, \mathbf{H}) + b \epsilon_b + b \cdot o(\log_2 P) \nonumber \\
& = & I(M_Y  ;  \mathbf{Y}_{[N_1'-l+1:N_1']} ~ \big| ~ M_Z, ~ \mathbf{H}) \nonumber  \\
& + & I(M_Y  ; \mathbf{Y}_{[1: N_1'-l]} ~ \big| ~ M_Z, ~ \mathbf{Y}_{[N_1'-l+1 : N_1']}, ~ \mathbf{H}) \nonumber \\
& + & b \epsilon_b + b \cdot o(\log_2 P). \label{eq: bound on d1 preliminary missing case general DoF}
\end{eqnarray}
The last equation is the counterpart of the bound (\ref{eq: bound on d1 preliminary missing case}) of Section \ref{sec: proof of thm: DoF region noCSIT IC missing case}.

Using the techniques developed in Lemma \ref{lem: main inequality missing case no CSIT DoF IC CRC}, it can be proved that
\begin{eqnarray*}
\lefteqn{ I(M_Y ;  \mathbf{Y}_{[1: N_1'-l]} ~ \big| ~ M_Z, \mathbf{Y}_{[N_1'-l+1 : N_1']}, \mathbf{H}) } \\
&& {} \leq \frac{N_1'-l}{M_2} I \big( M_Y  ;  \mathbf{Z}_{[1:M_2]} ~ \big| ~ M_Z, \mathbf{Z}_{[M_2+1:N_2]},  \mathbf{H} \big)  \\
&& {} + b \cdot o(\log_2 P) \\
&& {} \leq  \frac{N_1'-l}{M_2} \Big\{ I \big( M_Y, M_Z  ;  \mathbf{\tilde{Z}}_{[1:M_2]} ~ \big| ~ \mathbf{\tilde{Z}}_{[M_2+1:N_2]},  \mathbf{H} \big) \Big. \\
&& {} \Big. - b R_2 + b \epsilon_b \Big\} + b \cdot o(\log_2 P),
\end{eqnarray*}
where the last bound follows from (\ref{eq: tight bound on inter. at R2 IC missing case general DoF}). Substituting the above bound into (\ref{eq: bound on d1 preliminary missing case general DoF}), we obtain the following:
\begin{eqnarray*}
\lefteqn{ R_1 + \frac{N_1' - l}{M_2} R_2 = R_1 + \frac{N_1'-N_2 + M_2}{M_2} R_2 } \\
&& {} \hspace{-6mm} \leq \frac{1}{b} I(M_Y ~ ; ~ \mathbf{Y}_{[N_1'-l+1:N_1']} ~ \big| ~ M_Z, ~ \mathbf{H}) + \epsilon_b + 2 \cdot o(\log_2 P) \\
&& {} \hspace{-4mm} + \frac{N_1'-l}{M_2} \left\{ \frac{1}{b} I \big( M_Y, M_Z ;  \mathbf{\tilde{Z}}_{[1:M_2]}  \big|  \mathbf{\tilde{Z}}_{[M_2+1:N_2]},  \mathbf{H} \big) + \epsilon_b \right\} \\
&& {} \hspace{-8mm} \Rightarrow R_1 + \frac{N_1'-N_2 + M_2}{M_2} R_2\\
&& {} \hspace{-6mm} \leq \overline{\lim_{b \to \infty}} \left\{ \frac{1}{b} I(M_Y ~ ; ~ \mathbf{Y}_{[N_1'-l+1:N_1']} ~ \big| ~ M_Z, ~ \mathbf{H}) + o(\log_2 P) \right. \\
&& {} \hspace{-4mm} \left. + \frac{N_1'-l}{M_2} I \big( M_Y, M_Z ~ ; ~ \mathbf{\tilde{Z}}_{[1:M_2]} ~ \big| ~ \mathbf{\tilde{Z}}_{[M_2+1:N_2]}, ~ \mathbf{H} \big) \right\}.
\end{eqnarray*}
Since the last bound holds for all $(R_1,R_2) = \big( R_1(P),R_2(P) \big) \in \mathcal{C}(P)$, we have
\begin{eqnarray*}
\lefteqn{ d_1 + \frac{N_1'-N_2 + M_2}{M_2} d_2 \leq \overline{\lim_{P \to \infty}} ~ \frac{1}{\log_2 P} \Biggl\{  \Biggr. } \\
&& {} \hspace{-8mm} \overline{\lim_{b \to \infty}} \left\{ \frac{1}{b} I(M_Y ~ ; ~ \mathbf{Y}_{[N_1'-l+1:N_1']} ~ \big| ~ M_Z, ~ \mathbf{H}) + o(\log_2 P) \right. \\
&& {} \hspace{-8mm} \left. \left. + \frac{N_1'-l}{b M_2} I \big( M_Y, M_Z ~ ; ~ \mathbf{\tilde{Z}}_{[1:M_2]} ~ \big| ~ \mathbf{\tilde{Z}}_{[M_2+1:N_2]}, ~ \mathbf{H} \big) \right\} \right\},
\end{eqnarray*}
from which the desired bound can be derived by applying the single-user bound (the DoF of the point-to-point MIMO channel are limited by the number of receive antennas).

\section{Proof of Corollaries \ref{cor: structure of Ztilde CRC missing case}, \ref{cor: structure of Y_Si CRC missing case}, and \ref{cor: structure of overline overline X CRC-B missing case}} \label{app: proof of corollaries on structure CRC missing case}

\subsection{Proof of Corollary \ref{cor: structure of Ztilde CRC missing case}} \label{sub-app: proof of cor: structure of Ztilde CRC missing case}

The mutual information terms appearing in (\ref{eq: bound1 on Interf. at R2 tighter CRC(actually) missing case}) and (\ref{eq: bound2 on Interf. at R2 tighter CRC(actually) missing case}) are to be computed with $\mathbf{\tilde{Z}} = \big\{ \tilde{Z}(t) \big\}_{t=1}^b$ and $\tilde{Z}(t)$ defined via equation (\ref{eq: defn of Ztilde CRC missing case}), so that we may write $\tilde{Z}(t) =$
\[
\big( Q^{22}(t) \big)^* H^{21}(t) X^1(t) +  R^{22}(t) X^2(t) \big( Q^{22}(t) \big)^* + W'(t).
\]
Note now that since $H^{11}(t)$ and $W'(t)$ contain i.i.d. $\mathcal{C}\mathcal{N}(0,1)$ entries and since they are independent of $Q^{22}(t)$, we have
\[
\big( Q^{22}(t) \big)^* H^{21}(t) \sim H^{11}(t) \mbox { and } \big( Q^{22}(t) \big)^* W'(t) \sim W'(t),
\]
implying that
\[
\tilde{Z}(t) \sim H^{21}(t) X^1(t) + R^{22}(t) X^2(t) + W'(t),
\]
which in turn implies the corollary since the mutual information depends only the distribution of the relevant random variables.

\subsection{Proof of Corollary \ref{cor: structure of Y_Si CRC missing case}} \label{sub-app: proof of cor: structure of Y_Si CRC missing case}

If $p$ and $q$ are positive integers such that $p \leq q$ and $S_i^a = [p:q]$, and if $r$ is a positive integer such that $S_i^f = \{fi1, fi2, \cdots, fir\}$ \footnote{For $i = 0$, we follow a convention that $r = 0$ so that $S_0^f = \phi$.}, then we can write
\begin{eqnarray*}
Y_{S_i}(t) & \hspace{-6pt} = \hspace{-6pt} & \begin{bmatrix} Y_p(t) &  \cdots & Y_q(t) & Y_{fi1}(t) & \cdots & Y_{fir}(t) \end{bmatrix}^T  \\
& = & H^{11}_{S_i}(t) \hspace{1pt} X^1(t) + H^{12}_{S_i}(t) ~ \! X^2(t) + \begin{bmatrix} W_{[p:q]}(t) \\ w_{fi[1:r]}(t) \end{bmatrix},
\end{eqnarray*}
where
\begin{eqnarray*}
H^{11}_{S_i}(t) = \begin{bmatrix} H^{11}_{[p:q]}(t) \\ H^{11}_{fi[1:r]} \end{bmatrix} \mbox{ and } H^{12}_{S_i}(t) = \begin{bmatrix} H^{12}_{[p:q]}(t) \\ H^{12}_{fi[1:r]} \end{bmatrix}.
\end{eqnarray*}
The claims about the distributional properties of $H^{11}_{S_i}(t)$, $H^{12}_{S_i}(t)$, and $W_{S_i}(t)$ follow from their definitions.

\subsection{Proof of Corollary \ref{cor: structure of overline overline X CRC-B missing case}} \label{sub-app: proof of cor: structure of overline overline X CRC-B missing case}

In the following, we drop the time index $t$. All the analysis applies for any give $t$. Let $\overline{\overline{X}}_i$ denote the $i^{th}$ entry of $\overline{\overline{X}}$, which by definition, is given by $\overline{\overline{X}}_i =  e_i^* \overline{X}$.

Let
\[
E = \begin{bmatrix} e_1 & e_2 & \cdots & e_{N_1} \end{bmatrix}
\]
and
\[
\big( V \big)^* = \begin{bmatrix} v_1 & v_2 & \cdots & v_{M_1+M_2} \end{bmatrix}.
\]
Note that the columns $v_{M_1+1}$, $v_{M_1 + 2}$, $\cdots$, $v_{M_1+M_2}$ belong to the span of columns $e_{N_1-M_2+1}$, $e_{N_1-M_2 + 2}$, $\cdots$, $e_{N_1}$ by construction. Since $E$ is unitary, $e_i$ is orthogonal to $e_{N_1-M_2+1}$, $e_{N_1-M_2 + 2}$, $\cdots$, $e_{N_1}$ for any $i \in [1:N_1-M_2]$. These two facts imply that
\[
e_i^* v_j = 0 ~ \forall i \in [1:N_1-M_2], ~ \forall j \in [M_1+1:M_1+M_2].
\]
For any $i \in [1: N_1-M_2]$ this yields
\begin{eqnarray*}
\overline{\overline{X}}_i & = & e_i^* \overline{X} \\
& = & e_i^* \begin{bmatrix} v_1 & \cdots & v_{M_1} & v_{M_1+1} & \cdots & v_{M_1+M_2} \end{bmatrix} \begin{bmatrix} X^1 \\ X^2 \end{bmatrix} \\
& = & \begin{bmatrix} e_i^* v_1 & \cdots & e_i^* v_{M_1} & 0 & \cdots & 0 \end{bmatrix} \begin{bmatrix} X^1 \\ X^2 \end{bmatrix} \\
& = & \begin{bmatrix} e_i^* v_1 & \cdots & e_i^* v_{M_1} \end{bmatrix} X^1,
\end{eqnarray*}
where the first equality follows from the definition of $\overline{\overline{X}}$. The corollary now follows from the last equality by noting that the channel matrices and $X^1$ all are independent of $M_Z$.

\section{Proof of Lemma \ref{lem: full-rankness CRC thm2 missing case}} \label{app: proof of lem: full-rankness CRC thm2 missing case}

It is sufficient to prove that
\[
f \Big(  I \big(M_Y; ~ \tilde{Y}_{ \mathcal{P}^c(k)}(t) ~  \big| ~ M_Z, \mathbf{H}, \mathbf{H}_f, ~ \tilde{Y}_{\mathcal{P}(k)}(t) ~ \big) \Big) = 0 ~ \forall t.
\]
To this end, note that $\tilde{Y}_{\mathcal{P}(k)}(t)$ consists of $N_1 = M_1 + M_2$ channel outputs. Hence, the above equality holds if, using the channel outputs $\tilde{Y}_{\mathcal{P}(k)}(t)$, a noisy version of the channel inputs $X^1(t)$ and $X^2(t)$ can be constructed. Moreover, this can be done provided the channel matrix corresponding to the outputs $\tilde{Y}_{\mathcal{P}(k)}(t)$ is full rank with probability $1$. More precisely, if we write
\[
\tilde{Y}_{\mathcal{P}(k)}(t) = G_k(t) \begin{bmatrix} X^1(t) \\ X^2(t) \end{bmatrix} + \rm{ noise}
\]
for some $N_1 \times N_1$ matrix $G_k(t)$ (recall that $N_1 = M_1 + M_2$), then the desired equality holds, provided $G_k(t)$ is full rank with probability $1$. Thus, the goal of the remainder of this appendix is show that $G_k(t)$ is almost surely full rank. Toward this end, we prove that no row of $G_k(t)$ can be written as a linear combination of the remaining of its rows with some non-zero probability.

First, note from equation (\ref{eq: defn of Ytilde_Pk missing case CRC}) that  $\tilde{Y}_{\mathcal{P}(k)}(t)$ can be written in the following form:
\[
\tilde{Y}_{\mathcal{P}(k)}(t) = \begin{bmatrix}
\begin{bmatrix} \tilde{Y}_{S_1 k}(t) & \tilde{Y}_{S_2 k}(t) & \cdots & \tilde{Y}_{S_n k}(t) \end{bmatrix}^T \\
\tilde{Y}_{S_{n+1} [1:M_2]}(t) \\
\tilde{Y}_{S_{n+2} [1:M_2]}(t) \\
\vdots \\
\tilde{Y}_{S_{n+m} [1:M_2]}(t) \\
\tilde{Y}_{S_0 [1:N_2]}(t) \end{bmatrix},
\]
where for an integer $l$,
\[
\tilde{Y}_{S_j [1:l]}(t) = \begin{bmatrix} \tilde{Y}_{S_j 1}(t) \\ \tilde{Y}_{S_j 2}(t) \\ \vdots  \\ \tilde{Y}_{S_j l}(t) \end{bmatrix}
\]
and $\tilde{Y}_{S_i j}$ is defined via equation (\ref{eq: def Ytilde_Si CRC missing case}).

Recall from Corollary \ref{cor: structure of Ytilde_Si CRC missing case} that we have
\begin{eqnarray*}
\tilde{Y}_{S_i}(t) = H^{11}_{S_i}(t) \cdot X^1(t) + R^{12}_{S_i}(t) \cdot X^2(t) + W_{S_i}(t),
\end{eqnarray*}
where $H^{11}_{S_i}(t)$ is an i.i.d. Rayleigh-faded matrix that is independent of upper-triangular matrix $R^{12}_{S_i}(t)$, and moreover, these matrices are independent across $i$. As a result, $G_k(t)$ can be expressed in the following form (where we omit the time index $t$):
\[
G_k = \begin{bmatrix} \bar{G} & \begin{bmatrix}
\vspace{3pt} (R^{12}_{S_1})_k \\
\vspace{3pt} (R^{12}_{S_2})_k \\
\vspace{3pt} \vdots \\
\vspace{3pt} (R^{12}_{S_n})_k \\
\vspace{3pt} (R^{12}_{S_{n+1}})_{[1:M_2]} \\
\vspace{3pt} (R^{12}_{S_{n+2}})_{[1:M_2]} \\
\vspace{3pt} \vdots \\
\vspace{3pt} (R^{12}_{S_{n+m}})_{[1:M_2]} \\
\vspace{3pt} (R^{12}_{S_0})_{[1:N_2]} \end{bmatrix} \end{bmatrix},
\]
where $\bar{G}$ is $N_1 \times M_1$ i.i.d. Rayleigh-faded matrix, which is independent of $R^{12}_{S_i}$ $\forall$ $i$, and $(R^{12}_{S_i})_k$ denotes the $k^{th}$ row of $R^{12}_{S_i}$ while $(R^{12}_{S_i})_{[1:k]}$ denotes the matrix formed by retaining just the first $k$ rows of $R^{12}_{S_i}$. In other words, all entries of $G_k$ are independent of each other. Moreover, every entry of it, which is not surely zero, follows a continuous distribution.

Consider the $(N_1-N_2+1)^{th}$ row of $G_k$. Recalling that $n + m M_2 = N_1 - N_2$, we may write
\[
(G_k)_{N_1 - N_2 + 1} =  \begin{bmatrix} (\bar{G})_{N_1-N_2+1} & (R^{12}_{S_0})_1 \end{bmatrix}.
\]
Since $(R^{12}_{S_0})_1$ can not contain any entry which is zero (even, almost) surely, all entries of $(G_k)_{N_1 - N_2 + 1}$ are independent and each of them follows a continuous distribution. Hence, the probability that $(G_k)_{N_1 - N_2 + 1}$ belongs to any fixed $N_1-1$ dimensional subspace of the $N_1$-dimensional Euclidean space is zero.

Now, all rows of $G_k$, except its $(N_1-N_2+1)^{th}$ row, can together span at most an $(N_1 -1)$-dimensional subspace. Since all rows of $G_k$ are independent, the probability that the $(N_1-N_2+1)^{th}$ row of it lies within the span of the remaining of its rows is zero. This implies that no row of $G_k$ can be written as a linear combination of the remaining of its rows with some non-zero probability. Hence, $G_k$ is full rank with probability $1$.

\section{Proof of Lemma \ref{lem: eq: Step IV.4 CRC thm2 missing case}} \label{app: proof of lem: eq: Step IV.4 CRC thm2 missing case}

Since the channel matrices and additive noises are always taken to be i.i.d. across time, it is sufficient to prove that conditioned on
\[
\Big\{ M_Y, M_Z, H(t), H_f(t) \Big\},
\]
the joint distribution of
\[
\Big\{ \tilde{Y}_{S_i [1:k]}(t), ~ \tilde{Y}_{S_0 [k+1:N_2]}(t) \Big\}
\]
is identical to that of
\[
\Big\{ \tilde{Z}_{[1:k]}(t), ~ \tilde{Z}_{[k+1:N_2]}(t) \Big\}.
\]

Toward this end, first recall from Corollary \ref{cor: structure of Ytilde_Si CRC missing case} that $\tilde{Y}_{S_i}(t) \sim \tilde{Y}_{S_j}(t) \sim \tilde{Z}(t)$ for any $i \not= j$. Moreover, we may write
\begin{eqnarray*}
\tilde{Y}_{S_i}(t) = H^{11}_{S_i}(t) \cdot X^1(t) + R^{12}_{S_i}(t) \cdot X^2(t) + W_{S_i}(t),
\end{eqnarray*}
where $H^{11}_{S_i}(t)$ is i.i.d. Rayleigh-faded matrix that is independent of the upper-triangular matrix $R^{12}_{S_i}(t)$. Also $R^{12}_{S_i}(t) \sim R^{22}(t)$ $\forall$ $i,t$, where $R^{22}(t)$ is defined in Lemma \ref{lem: QR  decomposn missing case} and every entry of $R^{12}_{S_i}(t)$ is independent of all other entries of it.

Let $\Big( R^{12}_{S_i}(t) \Big)_{[n_1:n_2]}$ denote the matrix formed by retaining just the $n_1^{th}$ to $n_2^{th}$ rows of $R^{12}_{S_i}(t)$. Then, for any $k \in [1:M_2]$, we have
\[
R^{12}_{S_i}(t) \sim \begin{bmatrix} \Big( R^{12}_{S_i} (t) \Big)_{[1:k]} \\ \Big( R^{12}_{S_0} (t) \Big)_{[k+1:N_2]} \end{bmatrix}
\]
since matrices $R^{12}_{S_i}(t)$ are i.i.d. across $i$ and $t$; and every entry of $R^{12}_{S_i}(t)$ is independent of any other entry of it. This in turn implies that
\[
\tilde{Y}_{S_i}(t) \sim \begin{bmatrix} \vspace{4pt} \big( \tilde{Y}_{S_i}(t) \big)_{[1:k]} \\ \vspace{4pt} \big( \tilde{Y}_{S_0}(t) \big)_{[k+1:N_2]} \end{bmatrix}.
\]
The lemma now follows by noting that $\tilde{Y}_{S_i}(t) \sim \tilde{Z}(t)$.

\bibliographystyle{IEEEtran}
\bibliography{v6_IC_CRC_noCSIT_DoF}
\end{document}